%% file: WDM-LSS.mnrasFINAL.tex
\documentclass[useAMS,usenatbib,usedcolumn]{mn2e}
\pdfoutput=1
\usepackage{graphicx}
\usepackage{color}
\input{defs.tex}
\def\lesssim{\la}
\def\gtrsim{\ga}

\title[Nonlinear structure formation in WDM models]{Nonlinear evolution of cosmological structures in Warm Dark Matter models}
\author[Schneider, Smith, Macci\`o and Moore]{Aurel Schneider$^{1}$\thanks{Email: aurel@physik.uzh.ch}, Robert~E.~Smith$^{1,2}$, 
Andrea~V.~Macci\`o$^{3}$, and Ben Moore$^{1}$\\
{$^1$Institute for Theoretical Physics, University of Zurich, Zurich CH 8057}\\
{$^2$Argelander-Institute for Astronomy, Auf dem H\"ugel 71, D-53121 Bonn, Germany}\\
{$^3$Max Planck Institut f\"ur Astronomie, K\"onigsstuhl 17, D-69117 Heidelberg, Germany}\\
}

\begin{document}


\pagerange{\pageref{firstpage}--\pageref{lastpage}} \pubyear{2011}

\label{firstpage}
\maketitle


\begin{abstract}
  The dark energy dominated warm dark matter (WDM) model is a
  promising alternative cosmological scenario. We explore large-scale
  structure formation in this paradigm. We do this in two different
  ways: with the halo model approach and with the help of an ensemble
  of high resolution $N$-body simulations. Combining these
  quasi-independent approaches, leads to a physical understanding of
  the important processes which shape the formation of structures. We
  take a detailed look at the halo mass function, the concentrations
  and the linear halo bias of WDM. In all cases we find interesting
  deviations with respect to CDM. In particular, the
  concentration-mass relation displays a turnover for group scale dark
  matter haloes, for the case of WDM particles with masses of the
  order $m_{\rm WDM}\sim0.25 \keV$. This may be interpreted as a hint
  for top-down structure formation on small scales. We implement our
  results into the halo model and find much better agreement with
  simulations.  On small scales the WDM halo model now performs as
  well as its CDM counterpart.
\end{abstract}


\begin{keywords}
Cosmology: theory - large-scale structure of the Universe - dark matter
\end{keywords}


\section{Introduction}\label{sec:Intro}

Over the last decade the vacuum energy dominated cold dark matter
(hereafter $\Lambda$CDM) scenario, has emerged as a standard model for
cosmology.  This owes largely to the combination of information from
galaxy clustering surveys such as the 2dFGRS and SDSS with WMAP
measurements of the temperature anisotropies in the microwave
background
\citep{Coleetal2005short,Tegmarketal2006short,Komatsuetal2011short}.
However, the nature of the two dark components in the $\Lambda$CDM
model are still completely unknown and it is therefore important to
keep exploring alternative models and test their compatibility with
observations.

In the $\Lambda$CDM model the dark matter is assumed to be composed of
heavy, cold thermal relic particles that decoupled from normal matter
very early in the history of the Universe
\citep{Peebles1982,Blumenthaletal1984,KolbTurner1990,Jungmanetal1996}.
Whilst there is a large body of indirect astrophysical evidence that
strongly supports CDM, there are some hints that it has shortcomings.
Firstly, CDM galaxy haloes contain a huge number of subhaloes
\citep{Mooreetal1999c,DiemandKuhlen2008,Springeletal2008,Stadeletal2009},
while observations indicate that only relatively few satellite
galaxies exist around the Milky Way and M31
\citep{Mooreetal1999c,Klypinetal1999}. Secondly, the highest resolution halo simulations show that the
slope of the inner density profile decreases linearly at smaller radii
\citep{Navarroetal1997,Mooreetal1999c,Diemandetal2004,Springeletal2008,Stadeletal2009},
whereas the density profiles inferred from galaxy rotation curves are
significantly shallower \citep{Mooreetal1999c} \citep[and for recent
  studies see][and references there
  in]{Swatersetal2003,Saluccietal2007,deBloketal2008,Gentileetal2009}.
Thirdly, the observed number of dwarf galaxies in the voids appears to
be far smaller than expected from CDM
\citep{Peebles2001,Tikhonovetal2009,PeeblesNusser2010}. Another example
is the excess in the prediction of dwarf galaxy concentrations
\citep{Lovelletal2011}. Whilst, it has become clear that some of these
discrepancies might be resolved through an improved understanding of
galaxy formation, they have led some to consider changes to the
$\Lambda$CDM paradigm.

One possible solution might be warm dark matter (WDM)
\citep{BondSzalay1983,Bardeenetal1986,Bodeetal2001}. In this scenario,
the dark particle is considered to be lighter than its CDM
counterpart, and so remains relativistic longer and also retains a
thermal velocity.  Since WDM particles are collisionless and decouple
early, they may `free-stream' or diffuse out of perturbations whose
size is smaller than the Jeans' length\footnote{Although originally defined in the context of gas dynamics, the Jeans length can be generalized to collisionless systems by replacing the sound speed with the velocity dispersion. The reason for this tight analogy lies in the linearized equation of perturbations, which has the same structure for gas and collisionless fluids \citep[see][for more details]{Peebles1982}.} in the radiation dominated Universe \citep{KolbTurner1990}. This free-streaming of the WDM
particles acts to damp structure formation on small scales. Two
potential candidates are the sterile neutrino
\citep{DodelsonWidrow1994,ShaposhnikovTkachev2006}, and the gravitino
\citep{Ellisetal1984,Moroietal1993,Kawasakietal1997,Gorbunovetal2008},
both of which require extensions of the standard model of particle
physics.

Recent observational constraints have suggested that sterile neutrinos
can not be the dark matter: the Lyman alpha forest
\citep{Seljaketal2006b,Boyarskyetal2009a} and QSO lensing
\citep{MirandaMaccio2007} bounds are $m_{\nu_s}>8\keV$, whilst those
from the X-ray background are $m_{\nu_s}<4\keV$
\citep{Boyarskyetal2008}\footnote{Lower bounds on the mass of a fully
  thermalized WDM particle can be obtained using \Eqn{eq:massrescale}
  \citep[see][]{Vieletal2005}.}.  However, a more recent assessment
has suggested that a better motivated particle physics model based on
resonant production of the sterile neutrino, may evade these
constraints: the Lyman alpha forest bound is brought down to
$m_{\nu_s}\gtrsim 2\keV$ and the X-ray background is pushed to
$m_{\nu_s}<50\keV$ (for very low mixing angles)
\citep{Boyarskyetal2009b}. It therefore seems that additional,
independent methods for constraining the $\Lambda$WDM scenario would
be valuable.

In \citet{Markovicetal2010} and \citet{SmithMarkovic2011}, it was
proposed that the $\Lambda$WDM scenario could be tested through weak
lensing by large-scale structure. The advantage of such a probe is
that it is only sensitive to the total mass distribution projected
along the line of sight. However, to obtain constraints on the WDM
particle mass, an accurate model for the nonlinear matter clustering
is required. In these papers, an approach based on the halo model was
developed. Accurate predictions from this model require: detailed
knowledge of the abundance of dark matter haloes, their spatial
large-scale bias, and their density profiles. In these studies, it was
assumed that the semi-analytic methods, which were developed for CDM,
would also apply to WDM.
 
In this paper we perform a series of very high resolution CDM and WDM
$N$-body simulations with the specific aim of exploring the halo model
ingredients in the $\Lambda$WDM scenario.  Over the past decade, there
have been a limited number of numerical simulation studies of
nonlinear structure formation in the WDM model
\citep{Colombietal1996,Mooreetal1999c,Colinetal2000,WhiteCroft2000,Avila-Reeseetal2001,
  Bodeetal2001,Bullocketal2002,ZentnerBullock2003,Colinetal2008,Zavalaetal2009,MaccioFontanot2010,Lovelletal2011,Vieletal2011,Dunstanetal2011}. In
most of these previous studies, conclusions have been drawn from
object-by-object comparison of a relatively small number of haloes
simulated in boxes of typical size $L=25\Mpc$. In this work we are
more interested in the overall impact that the WDM hypothesis has on
the statistical properties of large-scale structures. We therefore
simulate boxes that are 10 times larger than have been typically
studied before, hence having roughly $\sim1000$ times larger sampling
volume. This means, that our conclusions will have greater statistical
weight, than those from previous studies. Furthermore, our results
should be less susceptible to finite volume effects, which can lead to
underestimates of the nonlinear growth.

The paper is structured as follows: In \S\ref{sec:WDMtheory} we
provide a brief overview of the salient features of linear theory
structure formation in the WDM model and we review the halo model
approach. In \S\ref{sec:Simulations} we describe the $N$-body
simulations. In \S\ref{sec:Ingredients} we explore the main
ingredients of the halo model: the halo mass function, bias and
density profiles. In \S\ref{sec:Comparison} we compare the halo model
predictions for the matter power with our measurements from the
simulations. Finally, in \S\ref{sec:Conclusion} we summarize our
findings.


\section{Theoretical Background}\label{sec:WDMtheory}

In this section we summarize the linear theory for WDM and the
nonlinear halo model in this framework.


\subsection{Linear theory evolution of WDM}

The physics of the free-streaming or diffusion of collisionless
particles out of dark matter perturbations has been discussed in
detail by \citet{BondSzalay1983}\footnote{For some recent theoretical
  treatments of WDM, also see \citet{Boyanovsky2010} and
  \citet{deVegaetal2010,deVegaetal2011}}. An estimate for the free-streaming length
can be obtained, by computing the comoving length scale that a
particle may travel up until matter-radiation equality ($t_{\rm
  EQ}$). At this point, the Jeans' length drops dramatically and
perturbations may collapse under gravity. A simple formula for this is
given by \citet{KolbTurner1990}:
\be \lambda_{\rm fs} = \int_{0}^{t_{\rm EQ}} \frac{v(t)dt}{a(t)} 
\approx  \int^{t_{\rm NR}}_{0}
\frac{cdt}{a(t)} + \int_{t_{\rm NR}}^{t_{\rm EQ}} \frac{v(t)dt}{a(t)} \ ,
\ee
where $t_{\rm NR}$ is the epoch when the WDM particles become
non-relativistic, which occurs when ${T_{\rm WDM}<m_{\rm
    WDM}c^2/3k_{\rm B}}$, where $T_{\rm WDM}$ and $m_{\rm WDM}$ are
the characteristic temperature and mass of the WDM particles. In the
relativistic case, the mean peculiar velocity of the particle is
simply $v(t)\sim c$. In the non-relativistic regime its momentum
simply redshifts with the expansion: $v\propto a(t)^{-1}$. This leads
to:
\be \lambda_{\rm fs} \approx  r_{\rm H}(t_{\rm NR})
\left[1+\frac{1}{2}\log \frac{t_{\rm EQ}}{t_{\rm NR}}\right]\ ,\ee
where $ r_{\rm H}(t_{\rm NR})$ is the comoving size of the horizon at
$t_{\rm NR}$. On inserting typical values for $t_{\rm NR}$ we find the
scaling:
\ba
\lambda_{\rm fs} \, & 
\approx \, & \,0.4 \, \left(\frac{m_{\rm WDM}}{\keV}\right)^{-4/3}
\left(\frac{\Omega_{\rm WDM} h^2}{0.135}\right)^{1/3} \left[\Mpc\right] \ .
\ea
However, the real situation is more complex than this, since
fluctuations inside the horizon grow logarithmically during radiation
domination via the Meszaros effect and free-streaming does not switch
off immediately after $t_{\rm EQ}$. To understand the collisionless
damping in more detail, one must numerically solve the coupled
Einstein-Boltzmann system of equations for the various species of
matter and radiation. Several fitting formulae for the WDM density
transfer function have been proposed
\citep{Bardeenetal1986,Bodeetal2001} and here we adopt the formula in
\citet{Vieletal2005}:
\be T_{\rm
  WDM}(k)=\left[\frac{P_{\rm lin}^{\rm WDM}}{P_{\rm lin}^{\rm
      CDM}}\right]^{1/2}=\left[1+(\alpha k)^{2 \mu}\right]^{-5/\mu},
\label{TFwdm}
\ee 
with $\mu=1.12$ as well as 
\be 
\alpha = 0.049 \left[\frac{m_{\rm WDM}}{\keV}\right]^{-1.11}
\left[\frac{\Omega_{\rm
      WDM}}{0.25}\right]^{0.11}\left[\frac{h}{0.7}\right]^{1.22}\rm
Mpc/h.  
\ee
Note that in the above we are assuming that the WDM particle is fully
thermalized. Following \citet{Vieletal2005}, the masses of sterile
neutrino WDM particles $m_{\nu_s}$ can be obtained from $m_{\rm WDM}$
through the formula:
\be 
m_{\nu_s}=4.43 \keV \left(\frac{m_{\rm WDM}}{1 \keV}\right)^{4/3}
\left(\frac{w_{\rm WDM}}{0.1225}\right)^{-1/3} \ .\label{eq:massrescale}
\ee

The characteristic length-scale $\alpha$ is related to the
free-streaming scale $\lambda_{\rm fs}$, and we shall therefore make
the definition that $\alpha\equiv\lambda_{\rm fs}^{\rm eff}$ is an
{\em effective} free-streaming length scale. The length-scale
$\lambda_{\rm fs}^{\rm eff}$ can be used to introduce the
`free-streaming' mass scale:
\be M_{\rm fs}=\frac{4\pi}{3}\overline{\rho}\left(\frac{\lambda^{\rm
    eff}_{\rm fs}}{2}\right)^3,  \ee
where $\bar\rho$ is the background density of the universe. This mass scale is important as it defines the scale below which
initial density perturbations are completely erased.

We can define yet another length scale: the `half-mode' length scale
$\lambda_{\rm hm}$.  This corresponds to the length scale at which the
amplitude of the WDM transfer function is reduced to 1/2. From
\Eqn{TFwdm} we find:
\be \lambda_{\rm hm}=2\pi\lambda^{\rm eff}_{\rm
  fs}\left(2^{\mu/5}-1\right)^{-1/2\mu} \approx 13.93 \lambda^{\rm eff}_{\rm
  fs}\ .  \ee
This length scale leads us to introduce another mass scale, the
half-mode mass scale:
\be M_{\rm hm}=\frac{4\pi}{3}\overline{\rho}\left(\frac{\lambda_{\rm
      hm}}{2}\right)^3 \approx 2.7\times 10^3 M_{\rm fs}\ .  \ee
This mass scale is where we expect the WDM to first affect the
properties of dark matter haloes. In passing, this partly explains the
claims made by \citet{SmithMarkovic2011}, that, for instance, the mass
function of haloes would be significantly suppressed on mass scales
$M\sim100M_{\rm fs}$.

\begin{figure}
  \centering{
    \includegraphics[width=8.6cm]{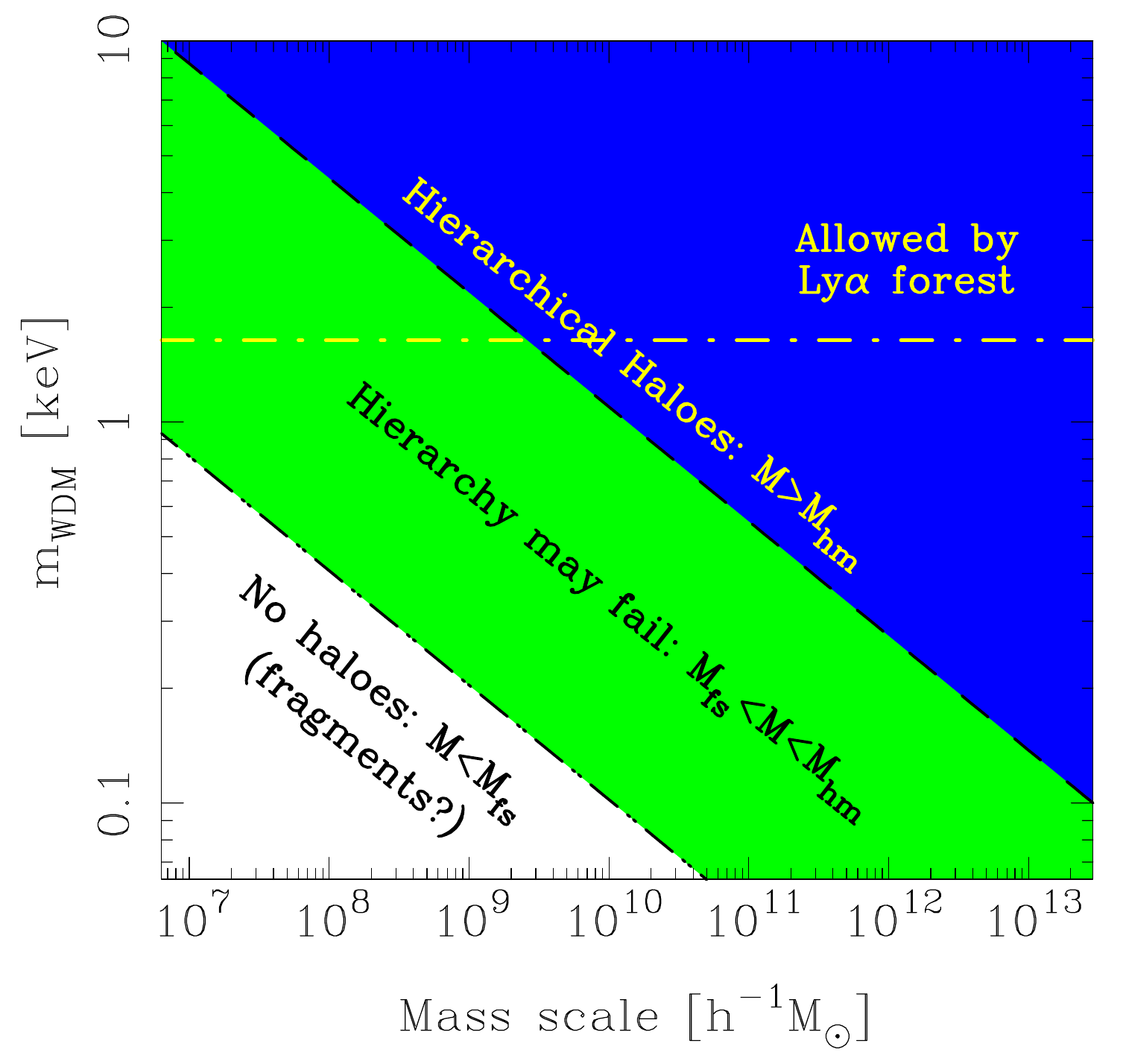}}
  \caption{\small{Free-streaming mass-scale ($M_{\rm fs}$) and
      half-mode mass scale ($M_{\rm hm}$) as a function of the mass of
      the WDM particle ($m_{\WDM}$). Haloes with masses $M>M_{\rm
        hm}$, may form hierarchically (upper right solid blue
      region). For haloes with masses $M_{\rm hm}>M>M_{\rm fs}$,
      hierarchical structure growth may fail (middle green
      region). For haloes with masses $M<M_{\rm fs}$, these may not
      form hierarchically since their initial peaks are completely
      erased (lower right empty region). However it is possible that
      such objects may emerge through fragmentation. The yellow
      dot-dash line denotes the current $m_{\WDM}$ allowed by the
      Lyman alpha forest \citep{Boyarskyetal2009a} (note that we have
      rescaled $m_{\nu_s}\rightarrow m_{\WDM}$ using
      \Eqn{eq:massrescale}).}}\label{fig:massscales}
\end{figure}

In \Fig{fig:massscales} we show the relation between $M_{\rm fs}$,
$M_{\rm hm}$ and the mass of the WDM particle candidate for our
adopted cosmological model. Three cases of relevance are apparent:
$M>M_{\rm hm}$, and haloes form hierarchically through accreting
material; $M_{\rm hm}>M>M_{\rm fs}$ and for these haloes the hierarchy
may fail with low mass haloes forming at the same time as higher mass
haloes; finally $M_{\rm fs}>M$ no halo formation, unless through the
fragmentation of larger structures. While the growth of overdensities is not affected above M$_{\rm hm}$, it is suppressed between M$_{\rm fs}$ and M$_{\rm hm}$, and should simply not take place below M$_{\rm fs}$.


\subsection{Nonlinear evolution of WDM: the halo model}\label{ssec:halomodel}

Cosmological structure formation is a very complicated, highly
nonlinear process that requires numerical simulation for a full
understanding. However, the halo model approach gives a simplified
analytical description of structure formation, which leads to
surprisingly good results \citep[and references
  therein]{CooraySheth2002}. Recently, the halo model has been adapted
for the WDM cosmological model by \citet{SmithMarkovic2011} and we now
summarize their basic approach.

The main idea of the halo model in WDM is to separate the density
field into a halo component, adding up all bound structure, and a
smooth component, standing for all matter, that has not collapsed due to free streaming.
This is different to the standard approach of the CDM halo model, where all matter is supposed to be in
bound structures.

Thus the WDM density field has the form,
\be
\rho(\mathbf{x})=\rho_{s}(\mathbf{x})+\sum_{i=1}^{N} M_i u(|x-x_i|,M_i)\ ,
\ee
where $\rho_s$ is the smooth part of the density field and
$u(x,M)=\rho_h(x|M)/M$ is the mass normalized density profile. The
average densities of the smooth and the bound components are then
given by
\be
\langle\rho\rangle = \overline{\rho} = 
\overline{\rho}_{\s} +\overline{\rho}_{\h},
\hspace{1cm}
\overline{\rho}_{\h} = f \overline{\rho}\ ,
\ee
where $f$ is the fraction of matter in bound objects. This can be
calculated by integrating over the halo mass function weighted by halo
mass:
\be\label{halofraction} f=\frac{1}{\overline{\rho}}\int_{0}^{\infty}
d\log M M \frac{d n}{d \log M} \ , \ee
where $dn=n(M)dM$ is the abundance of WDM haloes of mass $M$ in the
interval $dM$. The fraction $f$ is equal to unity in a perfectly hierarchical universe and drops below unity as soon as the mass function is suppressed due to the free streaming. In a WDM universe the amount of suppression depends on the mass of the WDM particle.

The power spectrum $P(k)$ is defined by the relation
\be \langle\delta(\mathbf{k})\delta(\mathbf{k'})\rangle \equiv
(2\pi)^3\delta_D(\mathbf{k}+\mathbf{k'})P(k)\ , \ee
where $\delta_D$ is the three dimensional Dirac delta function and
$\delta(k)$ is the Fourier transform of the matter overdensity
$\delta(\mathbf{x}) \equiv
(\rho(\mathbf{x})-\overline{\rho})/\overline{\rho}$. In terms of the
different density components, we can write:
\be
\delta(k)=f\delta_{\h}(k)+(1-f)\delta_{\s}(k),
\ee
where $\delta_{\chi} \equiv
(\rho_{\chi}-\overline{\rho}_{\chi})/\overline{\rho}_{\chi}$ with
$\chi\in\lbrace \h,\s\rbrace$. The power spectrum of the halo model
can now be determined by adding up the power spectra of the different
density components as well as their cross terms, giving
\be\label{hmPS}
P(k)=(1-f)^2P_{\s\s}(k) + 2(1-f)fP_{\s\h}(k) + f^2P_{\h\h}(k)\ .
\ee
The term $P_{\h\h}$ represents the power spectrum of matter trapped in
haloes, the term $P_{\s\s}$ designates the power spectrum of the
smooth component and the term $P_{\s\h}$ denotes the cross-power
spectrum between haloes and the smooth field.

The term $P_{\h\h}$ can be separated into one- and two-halo terms, which describe the power coming from the same halo, and the one coming from distinct haloes, respectively.
It can be expressed as:
\ba
P_{\h\h}(k) & = & P_{\h\h}^{2\h}(k)+P_{\h\h}^{1\h}(k) \ ;\label{PShh}\\ 
P^{\rm 2h}_{\h\h}(k) & = & \prod_{i=1}^{2}\left\{
\int_{0}^{\infty} \frac{dM_i}{\rhob_{\h}} M_i  n(M_i) {u}(M_i)\right\}\nonumber\\
&&\!\times P^{\rm c}_{\rm hh}(k|M_1,M_2), \  \label{PShh2}\\ 
P^{\rm 1h}_{\h\h}(k) & = &  
\frac{1}{\rhob_{\h}^2}\int_{0}^{\infty}dM n(M) M^2 {u}^2(k|M)\ ,\label{PShh1} 
\ea
where $u(k|M)$ is the Fourier transform of the mass normalized density
profile.
In \Eqn{PShh2} we have introduced the power spectrum of halo centers
$P_{c}^{\h\h}(k|M_1,M_2)$, which in general is a complicated function
of $k$ and the halo masses $M_1$ and $M_2$. However, if we neglect
halo exclusion and assume linear biasing with respect to the linear
mass density, then we may write this as,
\be\label{hcPS}
P^{c}_{\h\h}(k|M_1,M_2)\sim b_1(M_1)b_1(M_2)P_{\rm lin}(k)\ .
\ee
In this case, the function is separable and this considerably
simplifies the integrals in \Eqn{PShh2}. This approximation breaks
down on small, nonlinear scales, but on these scales, the two-halo
term is sub-dominant. The error induced by this approximation
(\ref{hcPS}) is most apparent at quasi-linear scales ($k\sim
[0.1,1.0]\kMpc$) and is $\lesssim30\%$. It is possible
to lower this error to $\lesssim5\%$ by using higher order
perturbation theory techniques and by including halo exclusion
\citep[see for example][]{Smithetal2011}. 
An easy but not fully consistent way of reducing the error down to $\lesssim10\%$ is to do the following replacement in Equation (\ref{hcPS}):
\be\label{replacement}
P_{\rm lin}(k) \rightarrow P_{\rm halofit}(k)W_{\rm TH}(kR),\hspace{0.5cm} R\simeq 2\Mpc,
\ee
where $W_{\rm TH}$ is the window function defined in \S\ref{sec:massfct} and $P_{\rm halofit}$ is the power spectrum calculated by the {\tt halofit} code \citep{Smithetal2003}.


The halo-smooth power spectrum is given by:
\be
P_{\s\h}(k)  =  \frac{1}{\rhob_{\h}}
\int dM n(M) M u(k|M) P^{c}_{\h\s}(k|M)\ ,
\ee
where $P^{c}_{\h\s}(k|M)$ is the power spectrum of the halo centers
with respect to the smooth mass field. On assuming that the smooth
field and the halo density field are linearly biased with respect to
the linear density field, we are lead to the relation:
\be\label{hcsP}
P^c_{\s\h}(k|M) \sim b_{s}b(M)P_{\rm lin}(k)\ ,
\ee
where $b_{s}$ is the linear bias of the smooth matter field defined in \S\ref{ssec:smoothbias}.
Finally, the smooth field auto-power spectrum is given by
\be\label{Pss}
P_{\s\s}(k) =  b_{\s}^2 P_{\rm lin}(k)\ . 
\ee
In order to reduce the error we can again replace the linear power spectrum in the Equations (\ref{hcsP}) and (\ref{Pss}), following the recipe of relation (\ref{replacement}).

On combining these power spectra, weighted by the correct functions of
their mass fractions, \`{a} la \Eqn{hmPS}, we find the total halo
model prediction for the nonlinear matter power spectrum in the WDM
model.


\begin{table*}
   \centering{
\begin{tabular}{|l|c|c|c|c|c|c|c|}
  \hline
  Sim label & $m_{\rm WDM}\, [\keV]$ & $M_{\rm fs}\, [\Msol]$ & $M_{\rm hm}
[\Msol]$  & $L\, [\Mpc]$
& $N_{\rm part}$ & $m_p \, [\Msol]$ & $l_{\rm soft} [\kpc]$ \\
  \hline
  CDM-S      &          &                  &                    &      &
$256^3$   &  $7.57 \times 10^{10}$ &  20\\
  CDM-M      & $\infty$ & 0                & 0                  &  256 &
$512^3$   &  $9.45 \times 10^{9}$  &  10\\
  CDM-L      &          &                  &                    &      &
$1024^3$  &  $1.18 \times 10^{9}$  &   5\\
  \hline
  WDM-1.25-S &          &                  &                    &      &
$256^3$   &  $7.57 \times 10^{10}$ &  20\\
  WDM-1.25-M & 1.25     & $2.3\times10^6$  & $6.3\times10^{9}$  &  256 &
$512^3$   &  $9.45 \times 10^{9}$  &  10\\
  WDM-1.25-L &          &                  &                    &      &
$1024^3$  &  $1.18 \times 10^{9}$  &   5\\
  \hline
  WDM-1.0-S  &          &                  &                    &      &
$256^3$   &  $7.57 \times 10^{10}$ &  20\\
  WDM-1.0-M  & 1.0      & $4.9\times10^6$  & $1.3\times10^{10}$  &  256 &
$512^3$   &  $9.45 \times 10^{9}$  &  10\\
  WDM-1.0-L  &          &                  &                    &      &
$1024^3$  &  $1.18 \times 10^{9}$  &  5\\
  \hline
  WDM-0.75-S &          &                  &                    &      &
$256^3$   &  $7.57 \times 10^{10}$ & 20\\
  WDM-0.75-M & 0.75     & $1.3\times10^7$  & $3.4\times10^{10}$  &  256 &
$512^3$   &  $9.45 \times 10^{9}$  &  10\\
  WDM-0.75-L &          &                  &                    &      &
$1024^3$  &  $1.18 \times 10^{9}$  & 5\\
  \hline
  WDM-0.5-S  &          &                  &                    &      &
$256^3$   &  $7.57 \times 10^{10}$ &  20\\
  WDM-0.5-M  & 0.5      & $4.9\times 10^7$ & $1.3\times 10^{11}$ &  256 &
$512^3$   &  $9.45 \times 10^{9}$  &  10\\
  WDM-0.5-L  &          &                  &                    &      &
$1024^3$  &  $1.18 \times 10^{9}$  &  5\\
  \hline
  WDM-0.25-S &          &                  &                    &      &
$256^3$   &  $7.57 \times 10^{10}$ &  20\\
  WDM-0.25-M & 0.25     & $5.0\times 10^8$ & $1.3\times 10^{12}$ &  256 &
$512^3$   &  $9.45 \times 10^{9}$  &  10\\
  WDM-0.25-L &          &                  &                    &      &
$1024^3$  &  $1.18 \times 10^{9}$  &  5\\
  \hline
\end{tabular}}
\caption{\small{WDM simulations.  From left to right, columns are:
      simulation name (S=Small, M=Medium, L=Large); mass of WDM
      particle ($m_{\rm WDM}$); free-streaming mass-scale $(M_{\rm
        fs})$; half-mode mass-scale ($M_{\rm hm}$); simulation
      box-size ($L$); number of particles ($N_{\rm part}$); mass of
      simulation particles $(m_{\rm p})$; comoving softening length
      ($l_{\rm soft}$).\label{simtable}}}
\end{table*}


\begin{figure}
\centering{ 
  \includegraphics[width=8.5cm]{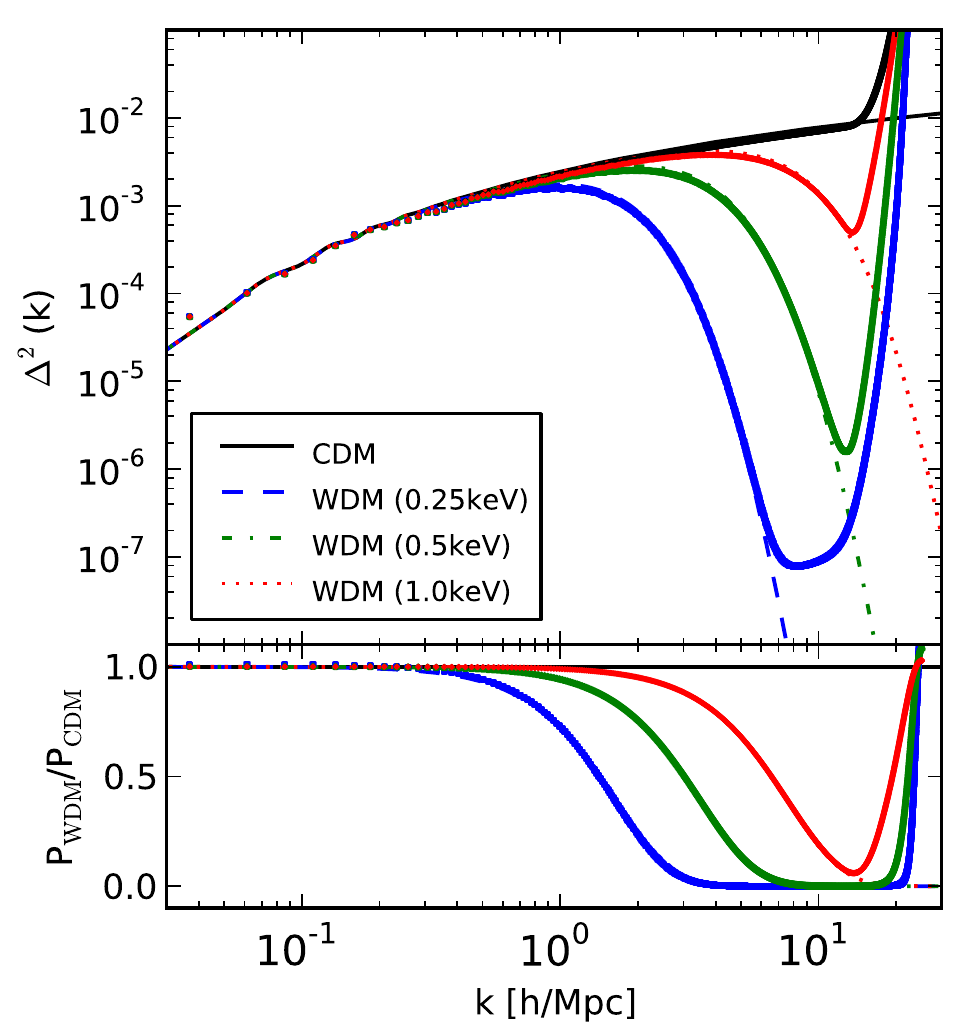}}
\caption{\small{Linear power spectra as a function of wavenumber in
    the CDM and WDM scenarios, at the initial redshift ($z=49$) of the
    simulations. {\em Top panel}: absolute dimensionless power:
    $\Delta^2=k^3P(k)/2\pi^2$. The lines 
    denote the linear power spectrum where
    $m_{\WDM}\in\{\infty,0.25,\,0.5,\,1.0\}\,{\rm keV}$.
    Points denote the power spectra measured from the initial
    conditions of the $N=1024^3$ simulations. {\em Bottom panel}:
    ratio of the initial WDM and CDM power spectra. Points and lines
    unchanged. }\label{linPowSpec}}
\end{figure}

 
\section{$N$-body Simulations of WDM}\label{sec:Simulations}

In order to study nonlinear structure growth in the WDM model, we have
generated a suite of $N$-body simulations. These were executed on the
{\tt zBOX3} supercomputer at the University of Z\"{u}rich. Each
simulation was performed using {\tt PKDGRAV}, a high order multipole
tree-code with adaptive time stepping \citep{Stadel2001}.

The cosmological parameters of the base $\Lambda$CDM model adopted,
are consistent with the WMAP7 best-fit parameters
\citep{Komatsuetal2011short} and we take: the energy-density
parameters in matter, vacuum energy and baryons to be $\Omega_{\rm
  m}=0.2726$, $\Omega_{\Lambda}=0.7274$, $\Omega_{\rm b}=0.046$; the
dimensionless Hubble parameter to be $h=0.704$; the primordial power
spectral index and present day normalization of fluctuations to be
$n_s=0.963$, and $\sigma_8=0.809$.

The CDM transfer function was generated using the code {\tt CAMB}
\citep{Lewisetal1999}. 
The linear power spectrum for each WDM model was then
obtained by multiplying the linear CDM power spectrum by $T_{\rm
  WDM}^2(k)$ from \Eqn{TFwdm}.  Initial conditions for each WDM model
were then generated at redshift $z=49$ using the serial version of the
publicly available {\tt 2LPT} code
\citep{Scoccimarro1998,Crocceetal2006}. In theory, we should also
include a velocity dispersion due to the fact that the particles still
retain a relic thermal velocity distribution. However, a quick
calculation of the rms dispersion velocity, showed that these effects
should be of marginal importance on scales $\gtrsim 50\kpc$ at the
initial redshift, and of order $\gtrsim 1\kpc$ at the present day for
$m_{\rm WDM}\ge 0.25\, [\keV]$. We therefore assume that their
inclusion will be a second order effect and so at this stage we
neglect them.

We generated initial conditions for a suite of simulations, one with a
CDM particle and five with WDM particle masses $m_{\rm WDM}\in \{0.25,
0.5, 0.75, 1.0, 1.25\} \keV$. For all runs, we set the box length
$L=256 \Mpc$.  This size is a compromise between choosing a box small
enough to accurately capture small-scale structure formation and large
enough to confidently follow the linear evolution of the box-scale
modes. This makes it possible for us to check agreement with the
linear theory and to measure linear bias.

Our simulations were also performed with three different mass
resolutions: $N=\{256^3,\,512^3,\,1024^3\}$. This enables us to
differentiate between genuine structures and spurious structures,
which can collapse out of the initial particle lattice
\citep[cf.][]{WangWhite2007,PolisenskyRicotti2010}. Full details of
the suite of simulations are summarized in Table~\ref{simtable}.

Dark matter haloes in the simulations were located using the
Friends-of-Friends algorithm \citep{Davisetal1985}. We used a modified
version of the {\tt skid} code, with the linking length parameter set
to the conventional value of $b=0.2$.

\Fig{linPowSpec} compares the initial linear theory power
spectra with the power spectra estimated from the initial conditions
of the $N$-body simulations, for the case \mbox{$N=1024^3$}. These
results show that, at the initial time, the WDM linear theory
distribution of fluctuations has been correctly seeded. It also shows a spike in the measured power spectrum at $k= 8\pi$ which is a consequence of the initial particle distribution on a grid.

\Fig{SimImages} presents a pictorial view of the growth of
structure in a selection of the simulations. The left column shows the
density evolution in a slice through one of the CDM simulations.  The
central and right panels show the same but for the case of WDM
particles with $m_{\rm WDM}=1.0\,\keV$ and $m_{\rm WDM}=0.25\,\keV$.
From top to bottom the panels show results for $z=4.4$, 1.1 and 0.


\input{images.tex}


\section{Halo Model ingredients in the WDM scenario}\label{sec:Ingredients}

In this section we detail the halo model ingredients and show how they
change in the presence of our benchmark set of WDM particle masses.


\begin{figure*}
  \centering{
    \includegraphics[width=12cm]{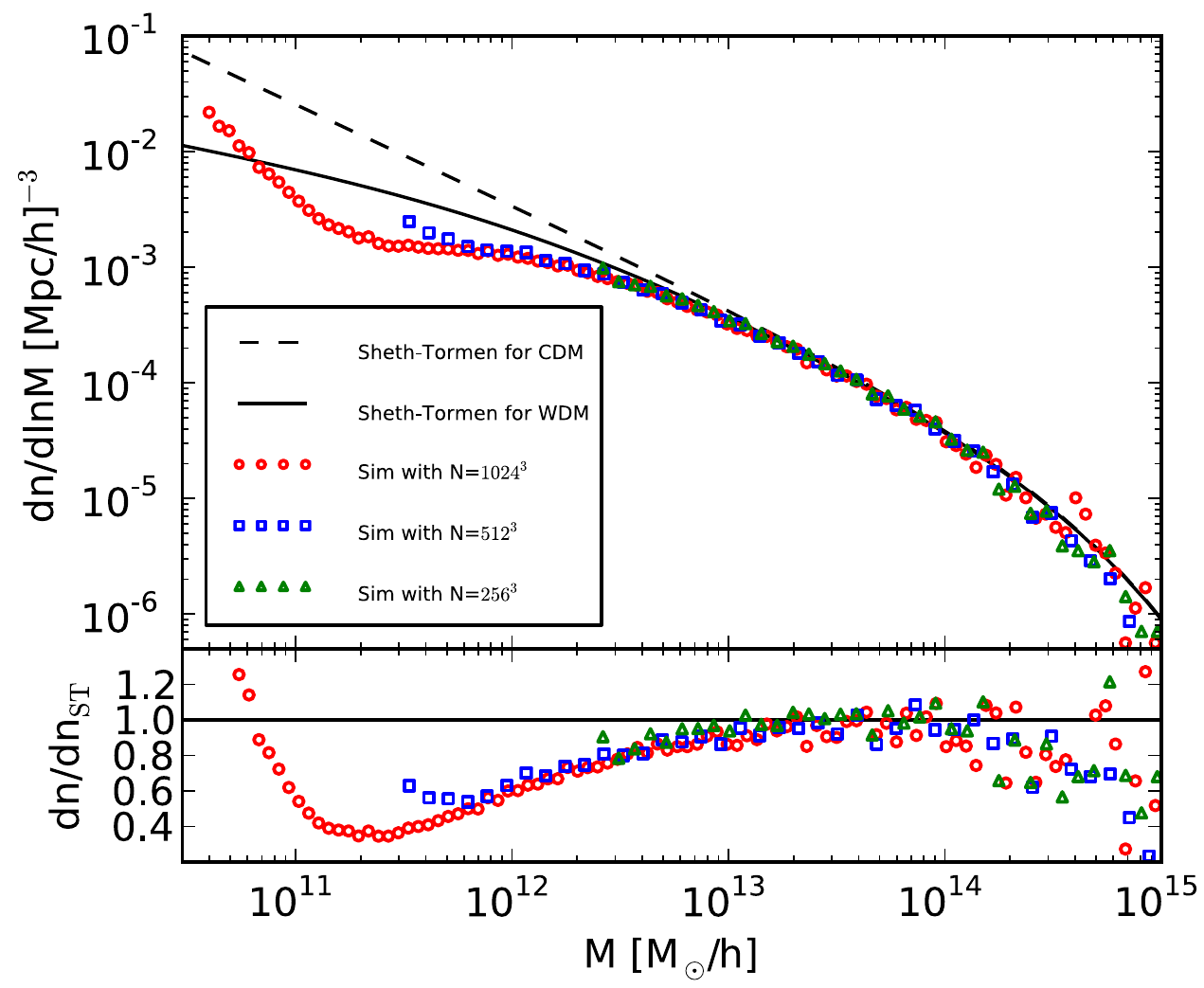}}
  \caption{\small{Measured mass function of the WDM simulations with
      $m_p=0.25$ keV and three different resolutions. The measurements
      lie below the Sheth-Tormen prediction, a well known result that
      is discussed in section \ref{sec:Ingredients}. The upturn of the
      mass function due to artificial haloes is visible in the
      simulations of high and medium resolution.\label{MassfctRes}}}
\end{figure*}


\subsection{Halo mass function}\label{sec:massfct}

In CDM the halo mass function can be explored through the excursion
set formalism \citep{PressSchechter1974,Bondetal1991}:
\be
\frac{dn}{d\log M}=-\frac{1}{2}\frac{\overline{\rho}}{M}
f(\nu)\frac{d\log\sigma^2}{d\log M} \ .
\ee
In the ellipsoidal collapse model of \citet{ShethTormen1999}, $f(\nu)$
is given by
\be f(\nu) = A\sqrt{\frac{2q\nu}{\pi}}
\left[1+(q\nu)^{-p}\right]e^{-q\nu/2},\hspace{0.4cm}
\nu=\frac{\delta_c^2(t)}{\sigma^2(M)} , \label{fnuST} \ee
with parameters: ${p=0.3}$, ${q=0.707}$ and with normalization
parameter ${A=0.3222}$. The linear theory collapse threshold is given
by, \mbox{$\delta_c(z)\equiv1.686/D(z)$}, where $D(z)$ is the linear
theory growth function. The variance on mass scale $M$ is,
\be \sigma^2(M) = \int \frac{\dk}{(2\pi)^3} P_{\rm Lin}(k) W_{\rm
  TH}^2(kR)\ , \ee
where $W_{\rm TH}(y) \equiv 3\left[\sin y - y \cos y\right]/y^3$ and
where the mass scale and radius of the filter function are related
through the relation: $M=4 \pi R^3 \rhob/3$.

The main idea in the excursion set approach is that there is a
monotonic mapping between the linear and nonlinear density
perturbations, averaged over a randomly selected patch of points in
the space. Further, the mapping can be calculated using the spherical
or ellipsoidal collapse approaches. The density perturbation in the
patch will collapse to form a virialized object when the linearly
extrapolated density in the patch reaches a certain collapse
threshold.  Despite the fact that this approach does not trace the
full complexity of nonlinear structure formation, the actual
predictions are in close agreement with measurements from
simulations. That is, at least for a CDM cosmology with well
calibrated values for the ellipsoidal parameters $p$ and $q$ and a
given halo finding algorithm. One important assumption, which is
implicit within this framework, is that structure formation must
proceed hierarchically.


\begin{figure*}
  \centering{
    \includegraphics[width=12cm]{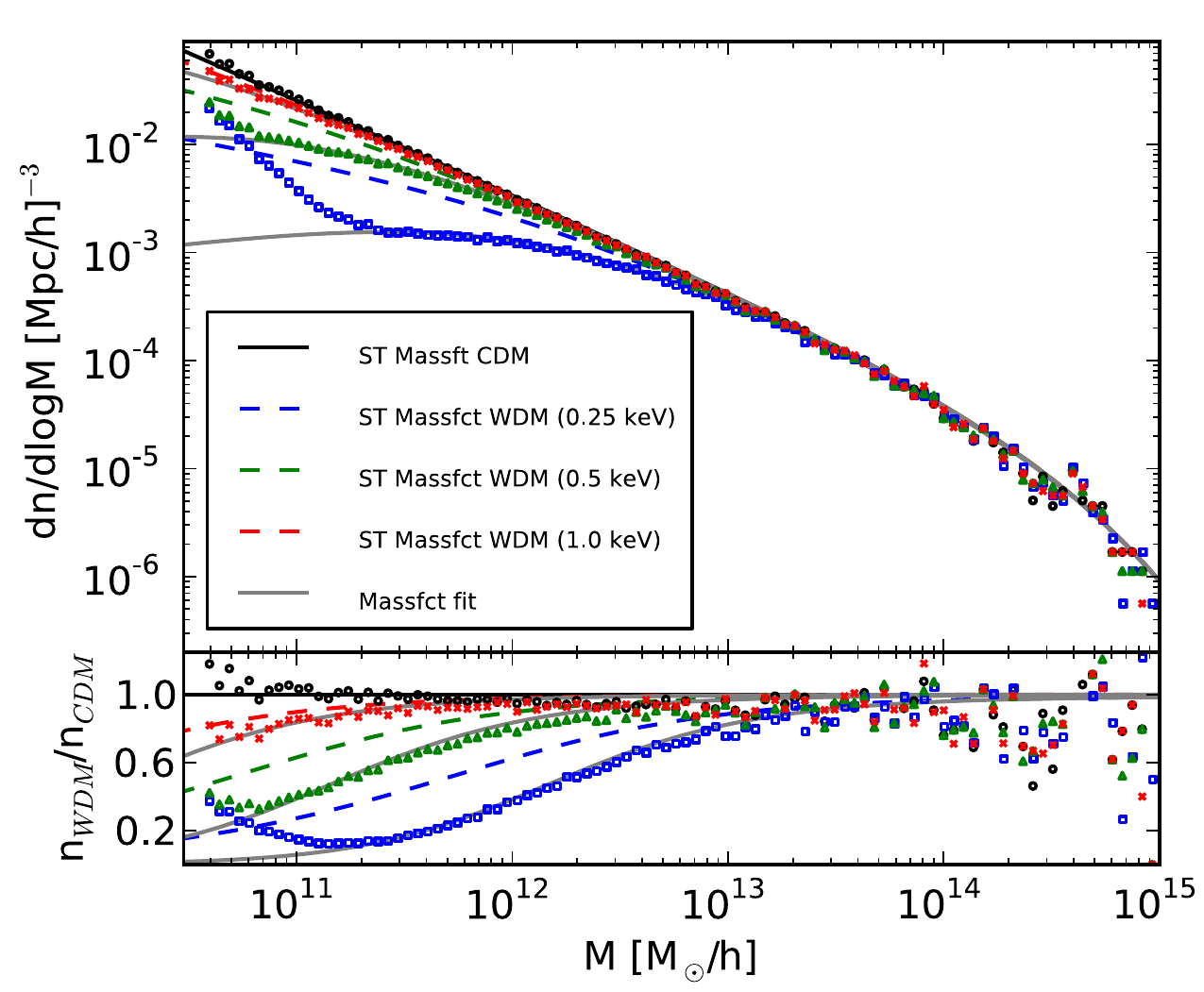}}
  \caption{\small{Comparison between the Sheth-Tormen mass functions
      (black solid line for CDM, colored dashed lines for WDM) and the
      measurements from the simulations (black circles for CDM,
      colored squares, triangles and crosses for WDM). The grey solid
      lines correspond to the mass function fit of
      \Eqn{MassfctFit}.\label{Massfct}}}
\end{figure*}


In the WDM scenario, things are more complicated, since structure
formation may not always proceed hierarchically. As described in
\S\ref{sec:WDMtheory}, we can identify three regimes of interest: for
$M>M_{\rm hm}$, the variance of WDM fluctuations becomes virtually
indistinguishable from that for CDM, and the excursion set approach
should be valid; for $M<M_{\rm fs}$ all primordial overdensities are
erased through diffusion of particles during the epoch of radiation
domination and we expect that no hierarchical halo formation will take
place on these mass scales. In between, where $M_{\rm hm}>M>M_{\rm
  fs}$, the WDM overdensity field is suppressed, but there is still
some power left that may enable hierarchical collapse to take
place. It is not clear {\em a priori}, how the mass function behaves
on these scales and whether the extended Press-Schechter approach
remains valid. We now investigate this using our simulations.

In \Fig{MassfctRes} we show the $z=0$ mass function of dark
matter haloes for the case of $m_{\rm WDM}=0.25\keV$. The figure
demonstrates the behaviour of the mass function as the simulation
resolution is increased from $N=256^3$, to $512^3$, to $1024^3$
particles, denoted by the triangles, squares and circles,
respectively. We can now see the effect of artificial clumping (cf.
discussion in \S\ref{sec:Simulations}), which is manifest as the
upturn of the curves at the low mass end of the mass function. One
common approach to dealing with this artificial clumping is to assume
that the simulations can be trusted down to the mass-scale just above
the up-turn. We also find, in agreement with \citet{WangWhite2007},
that this mass-scale increases as $N^{1/3}$, i.e. the inter-particle
spacing. In order to decrease the resolved mass by a factor of two, the particle resolution has to go up by a factor of eight. This is one of the main reasons why simulating WDM models is
significantly more challenging than simulating CDM models.

\Fig{MassfctRes} also shows the prediction of the halo mass
function for CDM and for this WDM model, from the ST mass function.
The figure clearly shows that the suppression of the ST model is not
sufficiently strong to describe the data. In addition to this the ST mass function is diverging towards small masses, while we expect a realistic mass function to drop to zero at latest below the free streaming scale.

\Fig{Massfct} compares the measurements of the WDM mass
functions from a selection of our highest resolution simulations with
the CDM case. We note that, whilst for the case of CDM the ST model is
in very good agreement with the data, the WDM data all lie below the
Sheth-Tormen prediction. That is, at least in the mass range above the
artificial upturn of the mass function.

\begin{figure}
  \centering{
    \includegraphics[width=8.5cm]{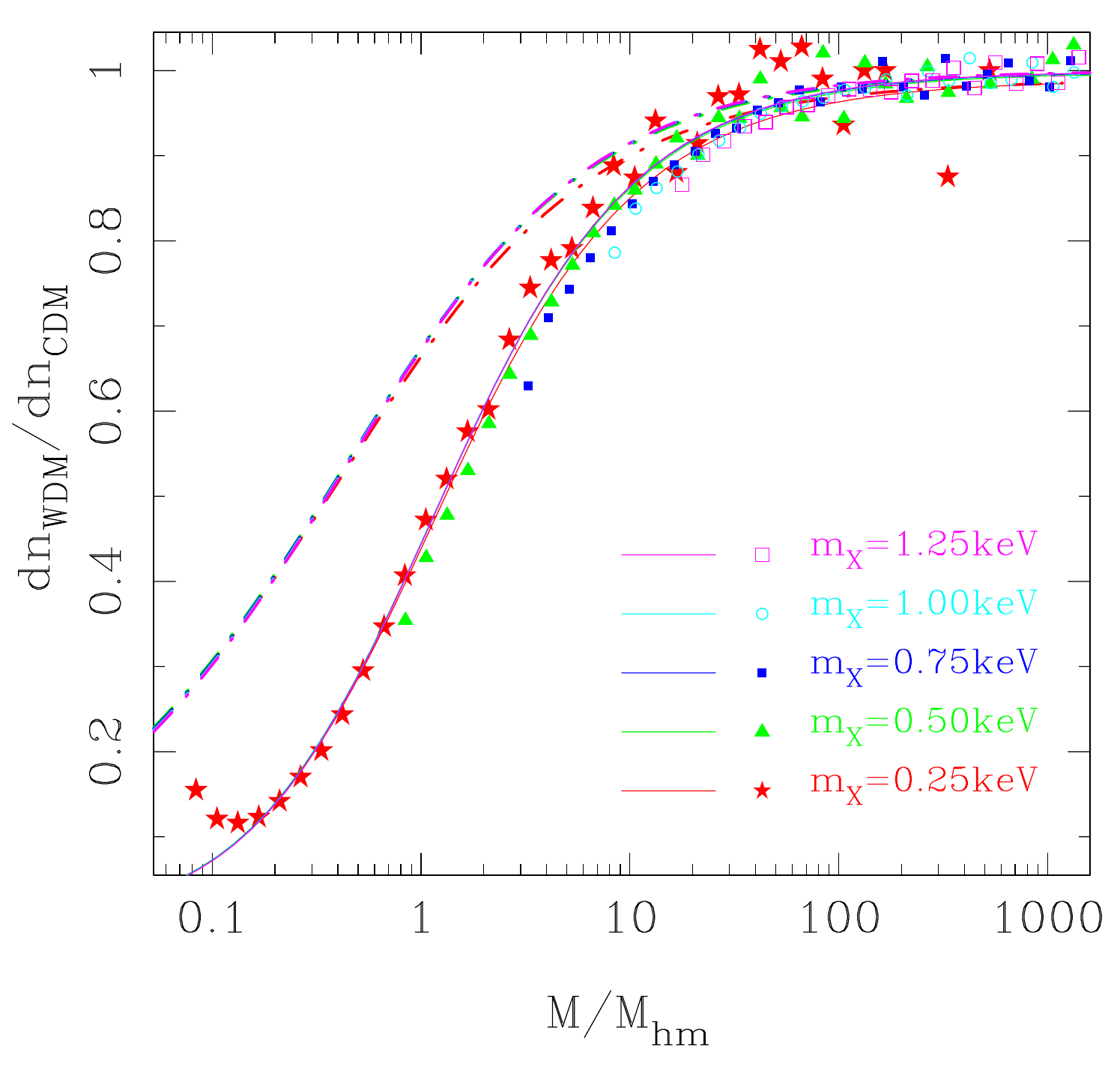}}
  \caption{\small{Ratio between the WDM and CDM mass
      functions, as a function of halo mass, scaled in units of the
      half-mode mass-scale $M_{\rm hm}$. The measurements from the
      $N=1024^3$ suite of WDM simulations are denoted by the point
      symbols. The dot-dashed lines denote the predictions from the
      \citet{ShethTormen1999} CDM mass function applied to WDM. The
      solid lines show the results from the fitting formula of
      \Eqn{MassfctFit}. \label{MassfctScale}}}
\end{figure}

Currently, there is no theoretical model that can explain the
discrepancy between the CDM and WDM measurements. We shall leave this
as an issue for future study. However, it is possible to develop a
fitting function that can describe the simulation results to high
accuracy. As first noted in \citet{SmithMarkovic2011}, if one rescales
the mass variable by $M_{\rm fs}$, or equivalently by $M_{\rm hm}$
(cf. \S\ref{sec:WDMtheory}), then the mass functions for a wide
variety of different values of $m_{\rm WDM}$ all appear to fall upon
the same locus\footnote{We find that the locus of theory curves is
  much tighter than was first noted in \citet{SmithMarkovic2011}. This
  owes to the fact that, they adopted the free-streaming scale of
  \citet{Bardeenetal1986,ZentnerBullock2003}, but used the transfer
  function of \citet{Vieletal2005} to generate the actual linear
  theory power spectra. This slight mismatch led to a slight off-set,
  which as \Fig{MassfctScale} shows, is removed when consistent
  definitions for $M_{\rm fs}$ and $M_{\rm hm}$ are adopted.}.


In \Fig{MassfctScale} we show that this scaling also works
surprisingly well for the mass function measured from the
simulations. We therefore look to fitting the rescaled mass
functions. After trying various forms, we found that the function
\be
\frac{n^{\rm A}_{\rm WDM}(M)}{n_{\rm WDM}^{\rm ST}(M)} 
= \left(1+M_{\rm hm}/M\right)^{-\alpha}, \label{MassfctFit}
\ee
which has only one free parameter $\alpha=0.6$ was able to fit all of
our data with an rms error well below five percent. Note that in the
above, $n_{\rm WDM}^{\rm ST}$ is the Sheth-Tormen model evaluated for
the WDM model in question. The resulting mass functions are plotted as
the grey solid lines in \Fig{Massfct}. A slightly worse fit may be
obtained by using the function,
\be 
\frac{n^{\rm B}_{\rm WDM}(M)}{n_{\rm CDM}(M)} = 
\left(1+M_{\rm hm}/M\right)^{-\beta}, \label{MassfctFit2} 
\ee 
with $\beta=1.16$ and this has the advantage that, one only needs to
evaluate the CDM mass function and rescale the masses.  We note that
whilst this paper was being prepared, a similar study was presented by
\citet{Dunstanetal2011}, who showed that $n^{\rm B}_{\rm WDM}$
provided a good description of their data but with the slightly higher
value $\beta=1.2$.

Finally, we examined the evolution of the WDM mass-functions up to
$z=1$ and found that \Eqn{MassfctFit} also provides a good description
of this data. The simplicity and generality of the fitting function
\Eqn{MassfctFit} is surprising and we think that it will be a useful
empirical formula.


\begin{figure}
  \centering{
    \includegraphics[width=8.7cm]{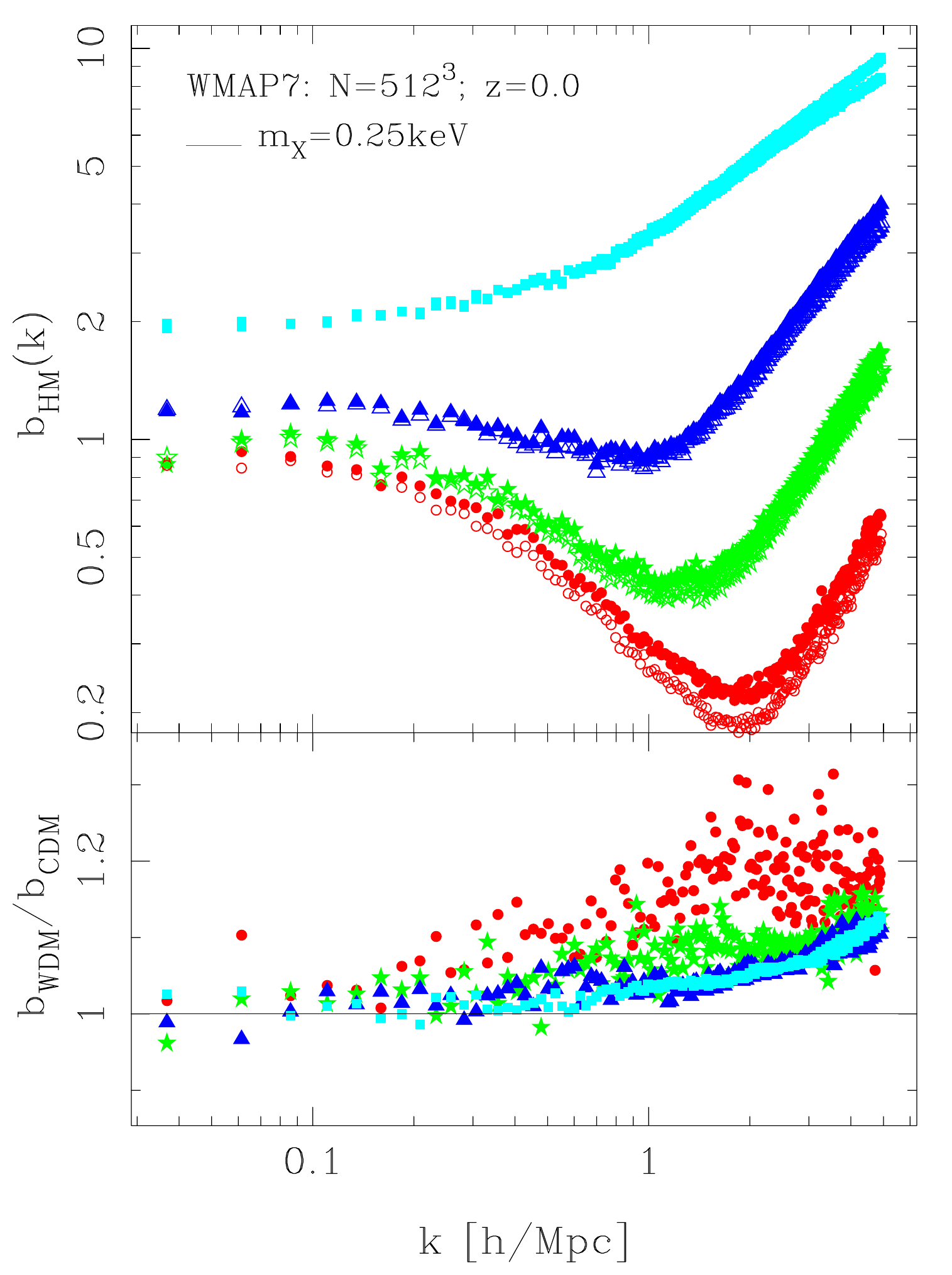}}
  \caption{\small{{\em Top panel}: Comparison of halo bias in the CDM
      and ${m_{\rm WDM}=0.25\keV}$ WDM model, as a function of wave
      number. The open and solid points denote the results for CDM and
      WDM, respectively. Circles, stars, triangles and squares denote
      results for haloes with masses in the range: $\log_{10}
      \left(M/[\Msol]\right)\in
      \{[12.0,12.5],[12.5,13.0],[13.0,13.5]{[>\!13.5]}\}$ {\em Lower
        panel:} Ratio of the bias in the WDM model with that for the
      CDM model. For $k>0.1\kMpc$ we see a relative excess signal in
      the bias of haloes with $M>10^{12}\Msol$ in the WDM models. For
      $k<0.1\kMpc$ the trends are unclear owing to sample variance.
      \label{scalebias}}}
\end{figure}


\begin{figure*}
\centering{
\includegraphics[width=10.2cm]{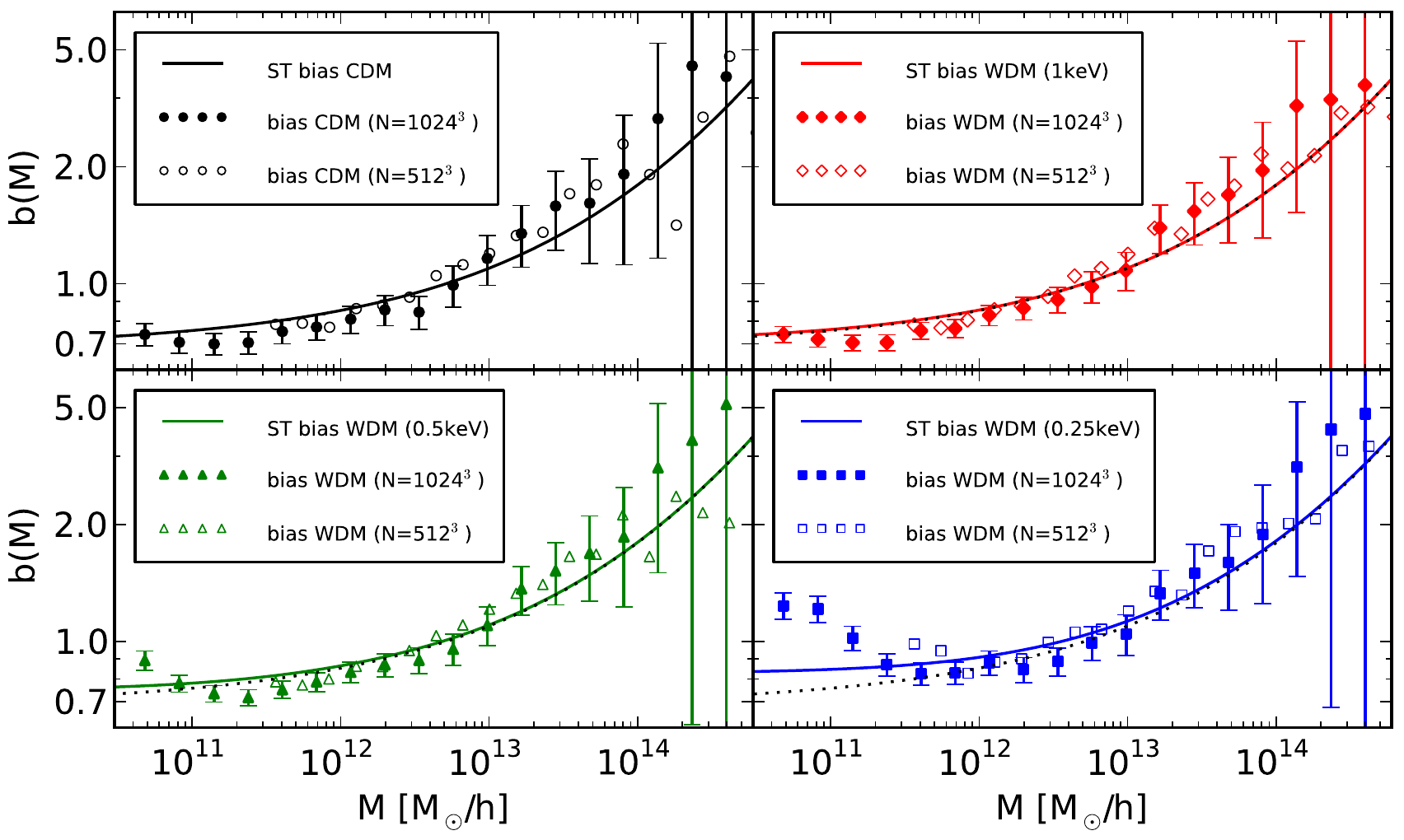}
\includegraphics[width=7.4cm]{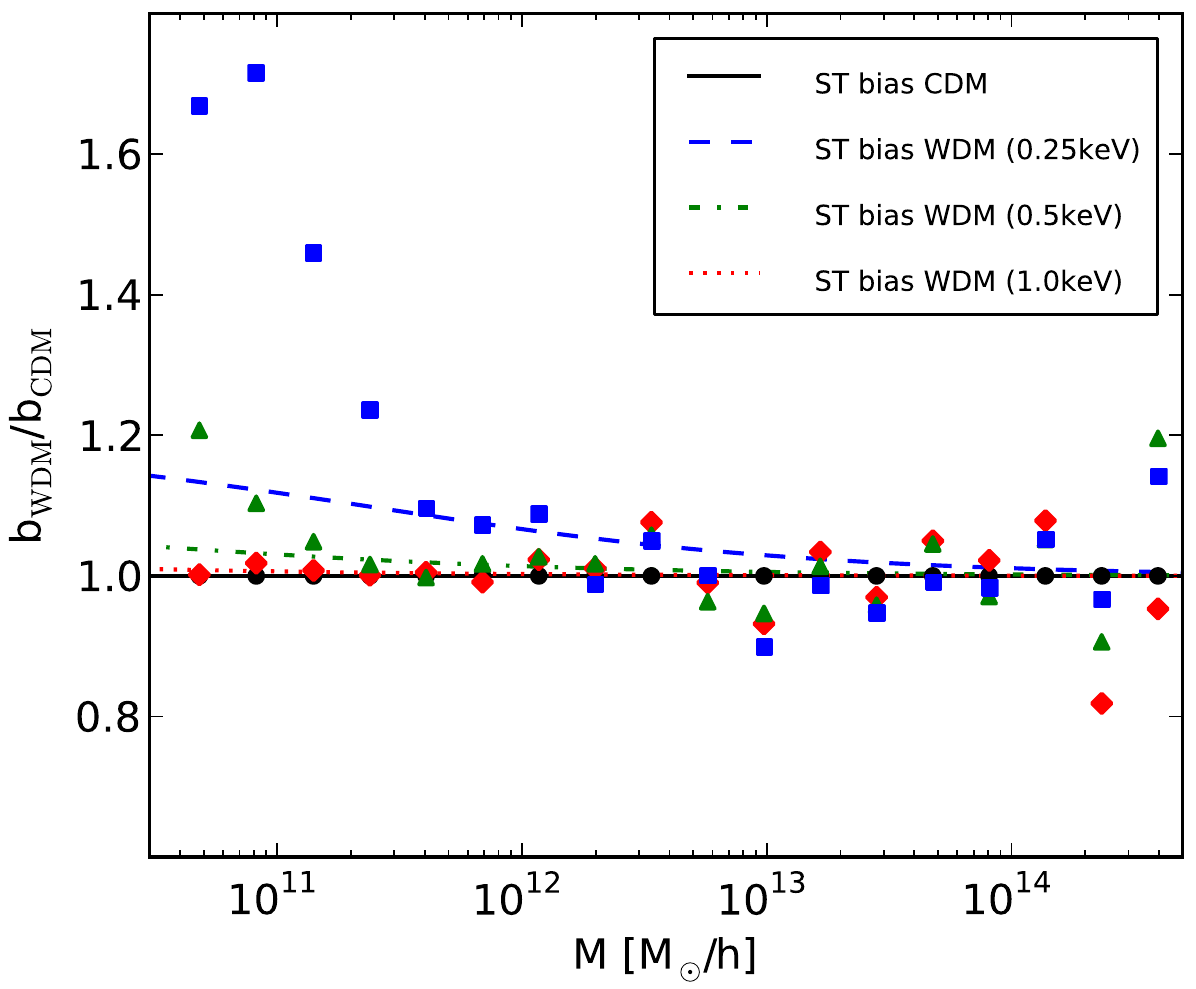}

}
\caption{\small{Left panel: Linear halo bias for CDM (top left) and WDM (top right: 1.0 keV, bottom left: 0.5 keV and bottom right: 0.25 keV). The filled and empty dots are measurements from the simulations with $N=1024^3$ and $N=512^3$, respectively. Error bars are calculated with an inverse variance wieghted estimator \citep[see][]{Smithetal2007}. The solid lines correspond to the Sheth-Tormen model prediction of Eq. (\ref{halobias}). The linear halo bias of CDM is shown as a black dotted line for comparison. Right panel: Ratios between the WDM and the CDM linear halo bias for the $N=1024^3$ runs. The error bars have been omitted for clarity. \label{bias}}}
\end{figure*}


\subsection{Halo bias}\label{ssec:halobias}

We are also interested in understanding how the density fields of dark
matter haloes and matter are related in the WDM framework. This
relation is usually termed bias, and if we assume that bias is local,
deterministic, and linear, then we may write:
\be \delta_{\rm h}(\bx|M) = b(M)\delta_{\rm m}(\bx)\ , \ee
where $b(M)$ is the linear bias coefficient, which depends only on the
mass of the halo. On using the excursion set formalism and the peak
background split argument, one may obtain a prediction for $b(M)$
\citep{ColeKaiser1989,MoWhite1996,ShethTormen1999}:
\be b^{\rm ST}(\nu)= 1 + \frac{q\nu
  -1}{\delta_c(z)}+\frac{2p}{\delta_c(z)\left[1+(q\nu)^p\right]} \ ,
\label{halobias}\ee
where the parameters $p$ and $q$ are as in \Eqn{fnuST}. As was shown
in \citet{SmithMarkovic2011}, if we apply this formula to the case of
WDM, then we would expect to see that for $M>M_{\rm hm}$ the bias
function is identical to that obtained for CDM. However, for $M<M_{\rm
  hm}$ we expect to find that the halo bias is increased relative to
the CDM case. 
This occurs due to the fact that $\nu$ tends towards a constant value for
$M<M_{\rm hm}$ and so $b^{\rm ST}$ becomes constant as well. We again use the
simulations to investigate these predictions.

In order to estimate the halo bias, we first sliced the halo
distribution into a set of equal number density mass bins. Then, for
each mass bin, we estimate the halo and matter auto-power spectra
$P^{\rm hh}(k|M)$ and $P^{\rm mm}(k)$, respectively. Our estimator for
the bias at each $k$-mode and in mass bin $M_{\alpha}$, can be
written:
\be
b_{i}(k_i,M_{\alpha})\equiv
\sqrt{\frac{P^{\rm hh}(k_i|M_{\alpha})-1/n_{\rm h}(M_\alpha)}
{P^{\rm mm}(k_i)}}\ ,
\ee
where $n_{\rm h}(M_\alpha)$ is the number density of haloes for the
mass bin $\alpha$.

\Fig{scalebias} compares the scale-dependence of the halo bias, for several mass bins, and as a
function of the wavemode, for the case of CDM (open points) and for
the $m_{\rm WDM}=0.25$ keV WDM model (solid points). Note that here we
actually present $b^{\rm hm}(k)\equiv P^{\rm  hm}/P^{\rm mm}$, where $P^{\rm hm}$ is defined by the relation $(2\pi)^3\delta_{\rm D}(\mathbf{k}+\mathbf{k'})P^{\rm hm}=\langle\delta_{\rm h}(\mathbf{k})\delta_{\rm m}(\mathbf{k'})\rangle$. In examining the ratio $b_{\rm
  WDM}/b_{\rm CDM}$, we see that there is increased bias in the WDM
case.

We then combine the estimates from each Fourier scale using a standard
inverse variance weighted estimator \citep[see
  e.g.][]{Smithetal2007}. Also, since, in this case, we are mainly
interested in determining the effective linear bias, we only include
modes with $k<0.1\kMpc$ (cf. \Fig{scalebias}). \Fig{bias}
presents the linear bias measurements together with the predictions
from $b^{\rm ST}(M)$ for a selection of the simulated WDM models. The
four panels show the cases: CDM, top left; $m_{\rm WDM}=1.0\keV$, top
right; $m_{\rm WDM}=0.5\keV$, bottom left; $m_{\rm WDM}=0.25\keV$,
bottom right. 

Considering the high mass haloes, we find that the bias estimates for
CDM and the WDM models appear to be in reasonable agreement with one
another. At lower masses, however, there is a prominent increase in the
bias for the WDM models with the lightest particle masses. We have
found that, rather than a genuine effect due to WDM initial
conditions, this boost appears to be a manifestation of the artificial
halo clumping discussed in \S\ref{sec:Simulations}. This becomes obvious by looking at the bottom-right plot in the panel on the left of \Fig{bias}, where the upturn of the high resolution simulation (solid blue squares) is shifted with respect to the upturn in the low resolution simulation (empty blue squares). The mass scales of the upturn in the halo bias coincides with the upturn in the mass function (see \Fig{MassfctRes}).
Importantly, this means that the halo-halo power spectrum
is strongly contaminated by the spurious
haloes, even on scales that are considered to be linear. To some
extent this is not so surprising, given that below a certain mass
scale we are dominated by spurious clumps.

In general, when considering masses above the minimum mass-scale that we
trust, rather than an excess bias with respect to $b^{\rm ST}$, we see
that the estimates appear to lie slightly below the theoretical
prediction at low masses. This has been observed in the CDM framework
by \citet{Tinkeretal2010}, and it seems to be the case for both our
CDM and WDM simulations. However, for the case of $m_{\rm WDM}=0.25$,
we do note that, just above the non-physical upturn, there is a sign
that bias in the WDM simulations is larger than in the CDM case.  This trend
is in qualitative agreement with the Sheth-Tormen prediction for
WDM. However, the effect is small and of the order of the error bars
and one would need both larger volume and higher resolution
simulations to robustly confirm this.


\begin{figure*}
  \centering{ 
    \includegraphics[width=13cm]{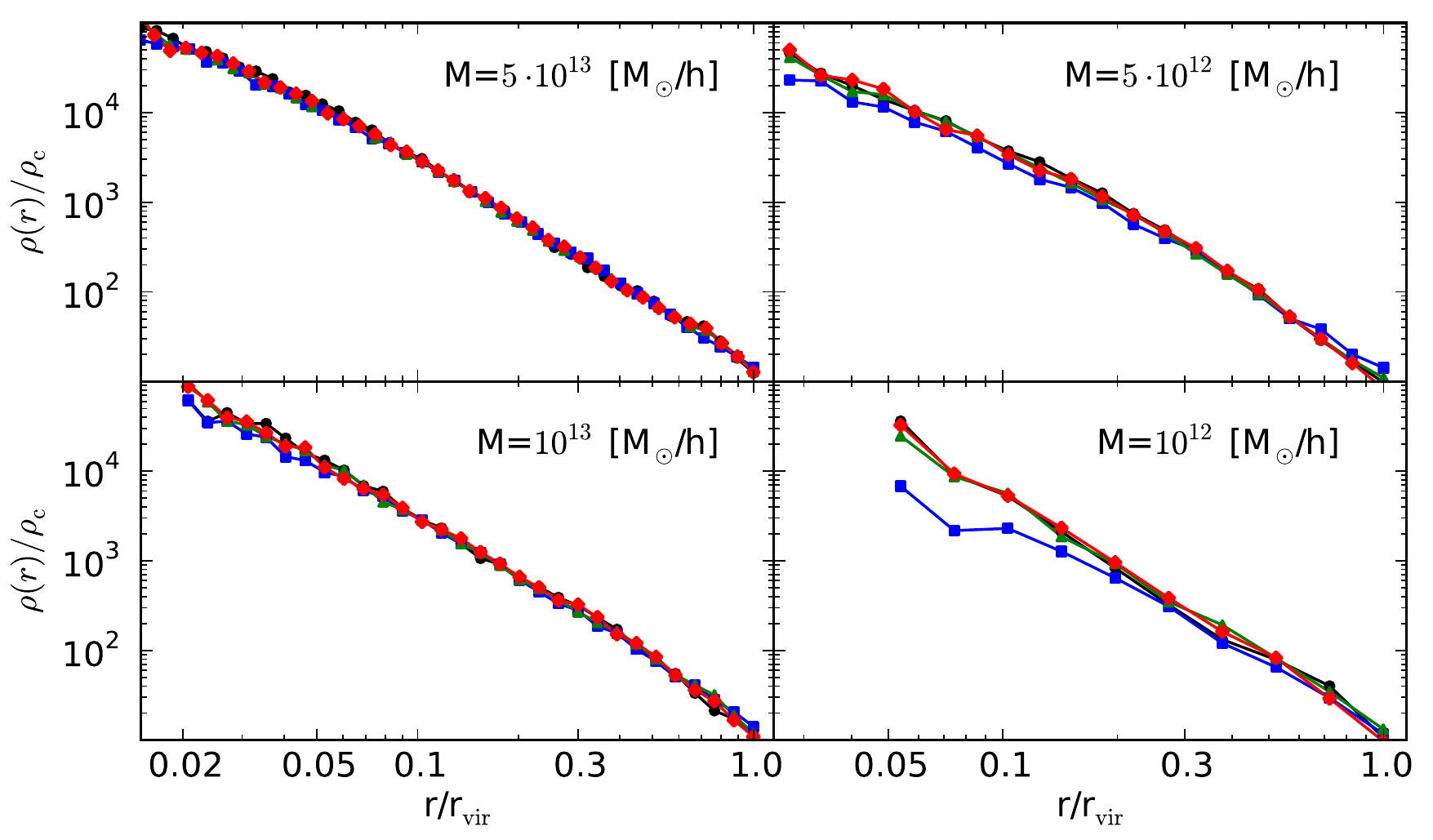}}
  \caption{\small{Measurement of the halo profiles for CDM (black) and
      WDM (blue: 0.25 keV, green: 0.5 keV, red: 1.0 keV) for different
      halo masses. The profiles of each mass bin are coming from a randomly chosen halo, which is identified in the CDM and all WDM simulations.}\label{profiles}}
\end{figure*}


\subsection{Anti-bias of the smooth component}\label{ssec:smoothbias}

We also require the density field of the smooth matter. As for the
case of the halo bias, if we assume that this is a linear,
deterministic function of the matter density, then we may write the
simple expression:
\be \delta_{\s}(\bx)=b_{\s}\delta(\bx) \ ,\ee
where $b_{\rm s}$ is the smooth bias parameter. As shown in
\citet{SmithMarkovic2011}, this can be calculated using a mass
conservation argument, and one finds:
\be
b_{s}=\frac{1}{1-f}\left[1-\frac{1}{\bar\rho}\int dM M n(M)b_1(M)\right] \le 1
\ee
Unlike the halo bias, which is mass dependent, the linear bias of the smooth
component stays constant over all scales. In consequence, the
smooth component of the power spectrum is directly proportional to the
linear matter power spectrum. We shall leave it for future study to
establish the veracity of this expression.


\subsection{Density profiles}\label{ssec:densityprofile}

Over the years, extensive numerical work has shown that, for the case
of the CDM model, the density profiles of dark matter haloes are
reasonably well characterized by the NFW profile
\citep{Navarroetal1997,Mooreetal1999c,Diemandetal2004,Springeletal2008,Stadeletal2009}. This has the universal form:
\be\label{NFWprofile}
\frac{\rho(r)}{\rhob} = \frac{\delta_{\rm s}}{y(1+y)^2} \ ; 
\hspace{0.5cm}y\equiv\frac{r}{r_{\rm s}}\ ,
\ee
where the two parameters $\delta_{\rm s}$ and $r_{\rm s}$ represent a
characteristic overdensity and scale radius.  The mass of each halo
can be determined by simply summing up the number of particles in a
given object and multiplying by the particle mass. We can connect this
to the virial radius through the relation:
\be\label{Mvir} M_{\rm vir}=\frac{4\pi}{3}\rhob \Delta_{\rm vir}r_{\rm
  vir}^3, \ee
where $r_{\rm vir}$ and $\Delta_{\rm vir}$ are the virial radius and
overdensity, respectively. The value of $\Delta_{\rm vir}$ is
typically chosen to denote the overdensity for virialization, and here
we adopt the value ${\Delta_{\rm vir}=200}$
\citep[e.g. see][]{ShethTormen1999}. However, the halo mass $M_{\rm
  vir}$ can also be obtained by integrating the density profile up to
$r_{\rm vir}$, which gives
\be\label{MvirINT}
M_{\rm vir}=4\pi \bar\rho  \delta_{\rm s}r_s^3 
\left[\log(1+c_{\rm vir})-c_{\rm vir}/(1+c_{\rm vir})\right] \ ,
\ee
where we have introduced the concentration parameter, defined as
$c_{\rm vir}\equiv r_{\rm vir}/r_{\rm s}$. On equating
\Eqns{Mvir}{MvirINT} we find that 
\be 
\delta_{\rm s} = \frac{c^3\Delta_{\rm vir}/3}
{\left[\log(1+c_{\rm vir})-c_{\rm vir}/(1+c_{\rm vir})\right]} \ .
\ee

This means that the original parameters $\{\delta_{\rm s},r_{\rm s}\}$
of the NFW profile can be replaced by $\{M_{\rm vir},c_{\rm
  vir}\}$. Thus, given a simulated halo of mass $M_{\rm vir}$, the
model has one free parameter, the concentration parameter $c(M)$

\Fig{profiles} shows the density profiles of several randomly
chosen haloes of different masses, for the case of CDM (black
connected points). We have matched these objects with their
counterpart haloes in the our standard set of WDM models and their
profiles are also plotted (coloured connected points). While these
profiles, on this logarithmic plot, all appear virtually
indistinguishable for high masses, there does appear to be a net
flattening off in the inner radius for the galaxy mass haloes in the
WDM model with $m_{\rm WDM}=0.25\keV$. One important point that can
not be easily gleaned from the figure, is that there is an overall
reduction in the masses of all the smaller haloes. As we will see,
this will have important consequences when we characterize the $c(M)$
relation in the next section. 

Earlier work on this topic by \citet{Mooreetal1999c} found that there
was almost no perceptible difference between CDM haloes and haloes
that formed from CDM initial conditions that had no small scale power
below a certain scale. Subsequent work by \citet{Avila-Reeseetal2001}
and \citet{Colinetal2008}, with a more careful treatment of the WDM
transfer function, have shown more significant differences. However,
in this case they were exploring models that were closer to HDM than
WDM. We therefore conclude that our results are broadly consistent
with all of these findings.

One further point, is that for this small sample, we see no visible
signs of the formation of a constant density core. This is in
agreement with work \citet{VillaescusaNavarroDalal2011}. Adding
thermal velocities into the simulations could in principle lead to the
formation of a constant density core through the
\citet{TremaineGunn1979} limit on the fine grained phase
space. However, the thermal velocities of WDM cool down with the
expansion of space and are already very small during the epoch of
structure formation. Thus, if cores are induced, we expect that they
will lie below the resolution limit of our simulations \citep[see][for
  more discussion of this]{KuziodeNarayetal2010}.

In summary, NFW profiles remain a valid approximation for density
profiles in our WDM simulations, given our spatial resolution and
choice of $m_{\rm WDM}$.


\begin{figure*}
\centering{
  \includegraphics[width=13cm]{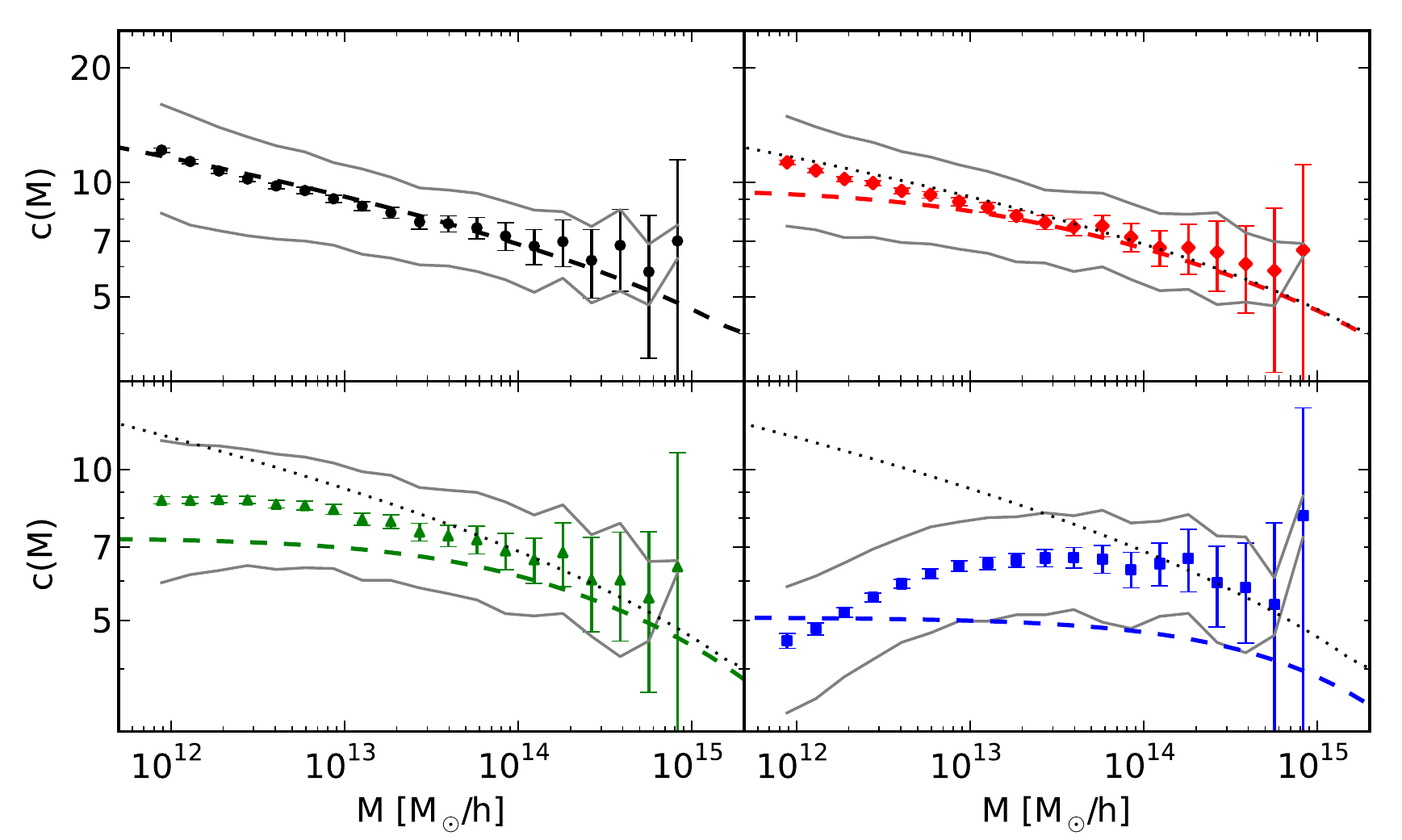}}
\caption{\small{Concentration to mass relation for CDM
    (top left) and for WDM with $m=1$ keV (top right), $m=0.5$ keV
    (bottom left) and $m=0.25$ keV (bottom right). The colored symbols
    denote the median concentrations, while the dashed
    lines correspond to the Bullock model with $F=0.001$ and $K=3.4$. For comparison the Bullock model for CDM has been added to the WDM plots in form of a black dotted line. The gray lines are the 1-$\sigma$ contours.\label{cFIT1}}}
\end{figure*}


\subsection{Concentration-mass relation}\label{ssec:concentrations}

As shown in the previous section the NFW model can be fully
characterized by specifying the concentration mass relation. We now
explore this for the case of WDM.

In the CDM model, $c_{\rm vir}$ has been shown to be a monotonically
decreasing function of $M_{\rm vir}$
\citep{Navarroetal1997,Bullocketal2001,Maccioetal2007,Maccioetal2008,Netoetal2007,Prada2011}. One
explanation for this, is that owing to the fact that haloes of
different mass form at different times, they are therefore exposed to
different background densities at collapse and this influences the
final core overdensity. A denser background during collapse leads to
generally higher concentrations. These ideas were encapsulated into a
simple model for halo concentration by \citet{Bullocketal2001}:
\be
c_{\rm vir}= K(z_{c}+1)/(z+1) \ ,
\ee
where $z_c$ is the redshift of collapse. This can be obtained by
solving the relation $\sigma(M_{*},z)=1.686$, where $M_*\equiv FM_{\rm
  vir}$, is defined to be a constant fraction of the virial mass. The
two constants $K$ and $F$ must be calibrated using numerical
simulations, and for our CDM simulations we found that $K=3.4$ and
$F=0.001$ provided a good fit to the data. However, we note that the
above arguments are only qualitatively correct, since, as first
pointed out by \citet{Bullocketal2001}, there is a large scatter
between $c_{\rm vir}$ and $M_{\rm vir}$. This can, in part, be traced
to the varying accretion histories and large-scale environments of
different haloes of the same final mass.

Turning to the WDM case, if we directly apply the Bullock model, but
using the WDM linear power spectrum, then we find a suppression and a
flattening of halo concentrations for masses $M<M_{\rm hm}$. Similar
to the mass function, this arises due to the fact that $\sigma(M)$ saturates to a constant value for masses approaching $M_{\rm fs}$. We have tested
these predictions, by estimating the concentration parameters for all
{\em relaxed} haloes in our CDM and WDM simulations that contain more
than $N=500$ particles \citep[for full details of the method that we
  employ see][]{Maccioetal2007,Maccioetal2008}.

\Fig{cFIT1} shows the measured halo concentrations as a
function of mass for a selection of the highest resolution CDM and WDM
simulations.
The gray solid lines correspond to the 1-$\sigma$ contours of the measurements, indicating a considerable spread in the concentration-mass relation. The large solid symbols denote the median, with the errors being computed on the mean, i.e. we use
$\sigma/\sqrt{N_i}$, where $N_i$ are the number of haloes in the $i$th
mass bin. The dashed lines denote the predictions from the Bullock
model. For the CDM case it works reasonably well, especially with our
modified parameters $\{K,F\}$. However, the model shows the wrong
qualitative behavior for the WDM scenario: whilst the curve for the
Bullock model always flattens out towards low masses, we see that for
the cases of the lighter WDM particles, there is a turnover in the
relation. This turnover in the \mbox{$c_{\rm vir}$--$M$} relation at
low masses is important, as it indicates the end of hierarchical
collapse and the emergence of a period of top-down structure formation. As a test of these results we performed additional WDM runs with the {\tt Gadget-2} gravity code \citep{Springel2005} and we observe the same turnover in this independent set of simulations.

In order to model the \mbox{$c_{\rm vir}$--$M$} relation for WDM, we
shall adapt the Bullock model. As in the case of the mass function, we
do this by introducing a correction function described by the
relation:
\be
\frac{c_{\rm WDM}(M)}{c_{\rm CDM}(M)}=
\left(1+\gamma_1\frac{M_{\rm hm}}{M}\right)^{-\gamma_2}\ ,
\label{cMfitfct}
\ee
where we have again rescaled the halo mass by $M_{\rm
  hm}$. Least-squares optimization of the free-parameters gives:
$\gamma_1=15$ and $\gamma_2=0.3$. 

\begin{figure}
\centering{ \includegraphics[width=8.4cm]{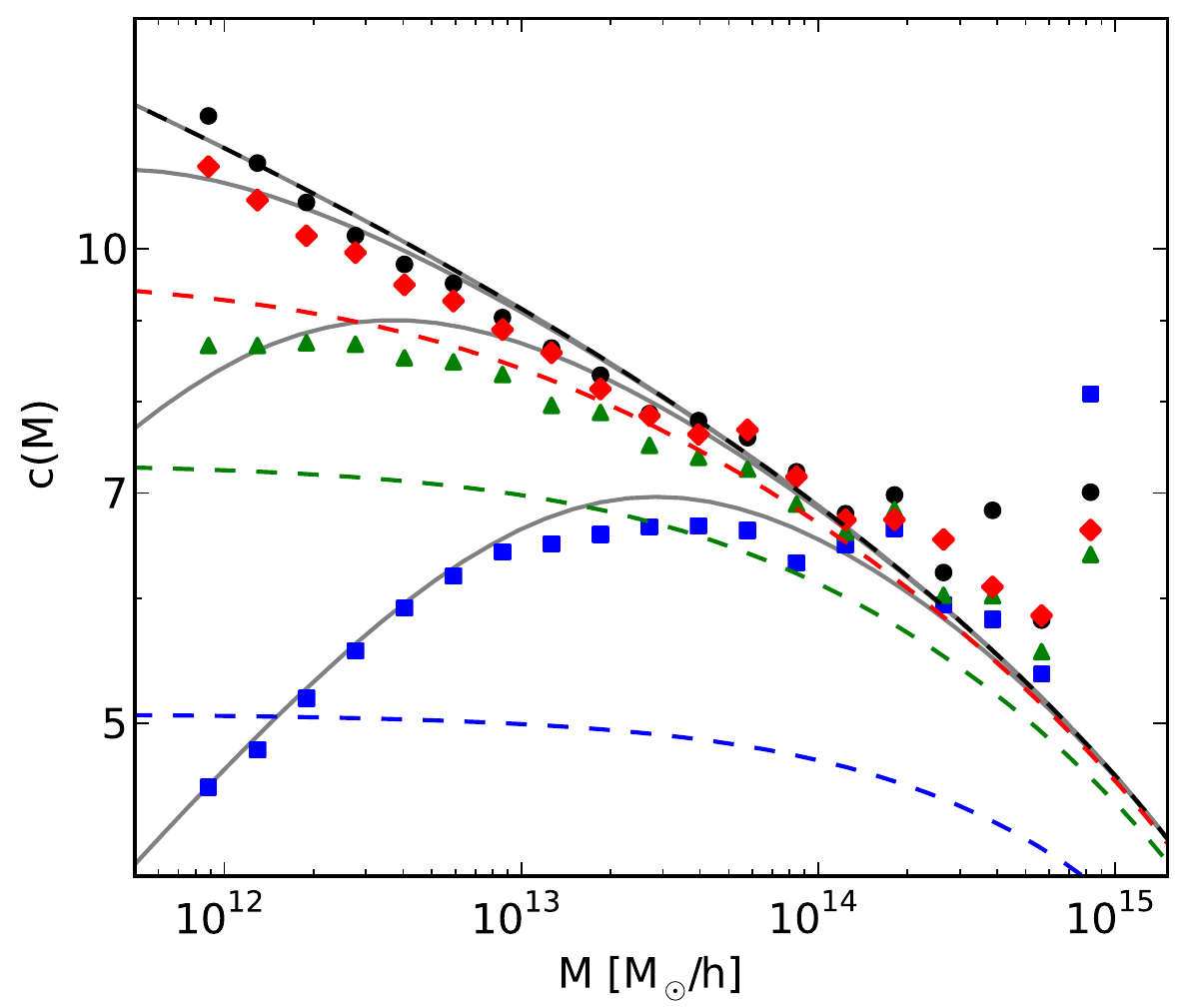}}
\caption{\small{Same as \Fig{cFIT1} but measurements are superimposed
    on one another and without error bars. The additional gray lines
    illustrate the fitting function from
    \Eqn{cMfitfct}.\label{cFIT2}}}
\end{figure}

In \Fig{cFIT2} we compare the fitting function (gray solid
lines) with the results from the simulations. The parametric relation
describes the $c_{\rm WDM}$--$M$ relation with a precision of
better than 10\% (the fit appears less satisfying for the case $m_{\rm
  WDM}=0.5 \keV$, but only for the lower mass bins). Interestingly,
the value $\gamma_1\sim10$, informs us that the $c(M)$ relation is
sensitive to the presence of WDM for mass scales one order of
magnitude larger than for the mass function.  As we will see in the
next section, this will be important for modeling the nonlinear power
spectrum on small scales.


\begin{figure}
  \centering{
    \includegraphics[width=8.6cm]{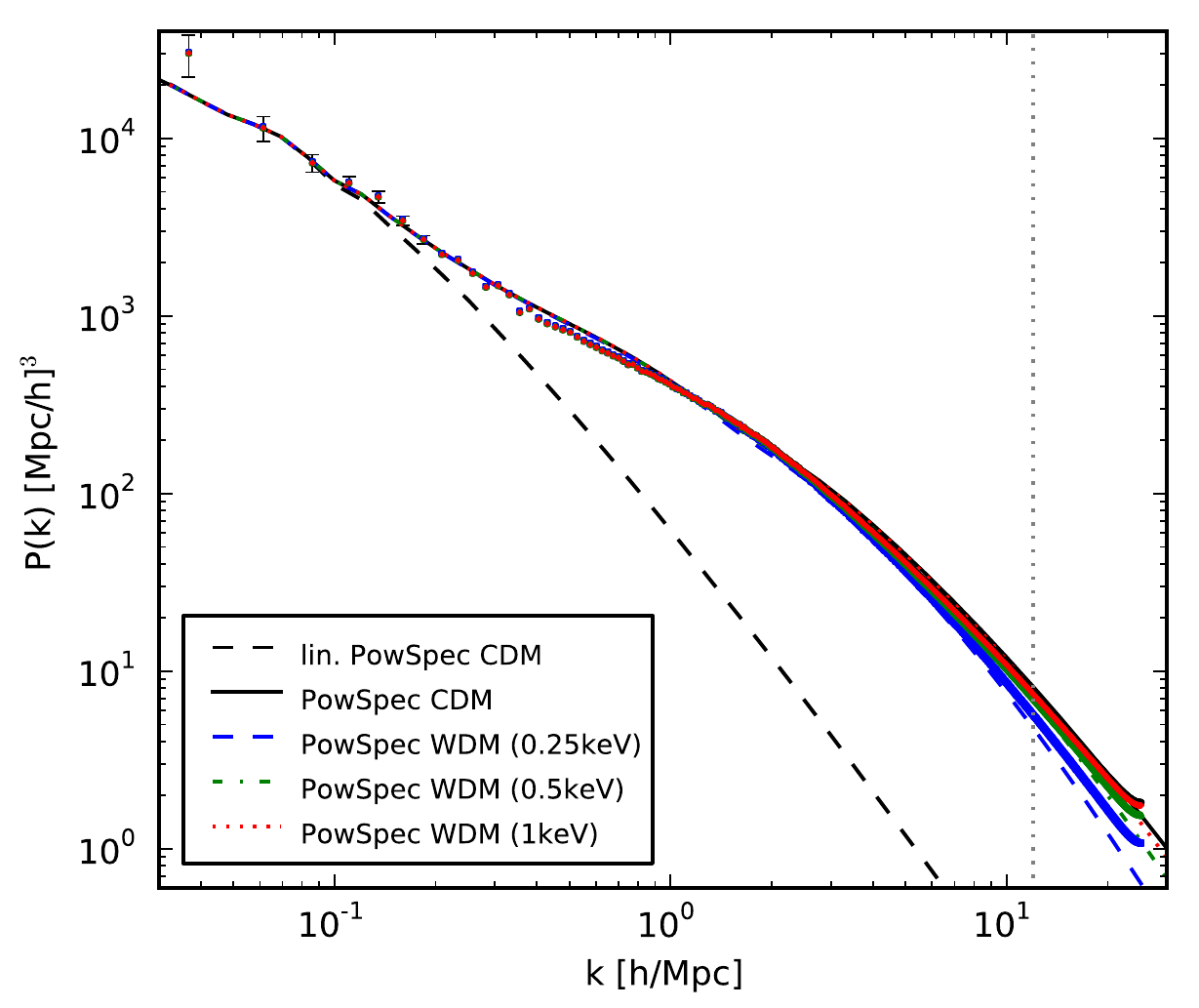}}
  \caption{\small{Nonlinear power spectra from the simulations (dots)
      and from the original halo model (lines), developed by
      \citet{SmithMarkovic2011}. Black corresponds to CDM and color to
      WDM (red: 1 keV, green: 0.5 keV, blue: 0.25 keV). The vertical
      gray dots indicate half the Nyquist frequency.}\label{PScomparison0}}
  \centering{    
    \includegraphics[width=8.55cm]{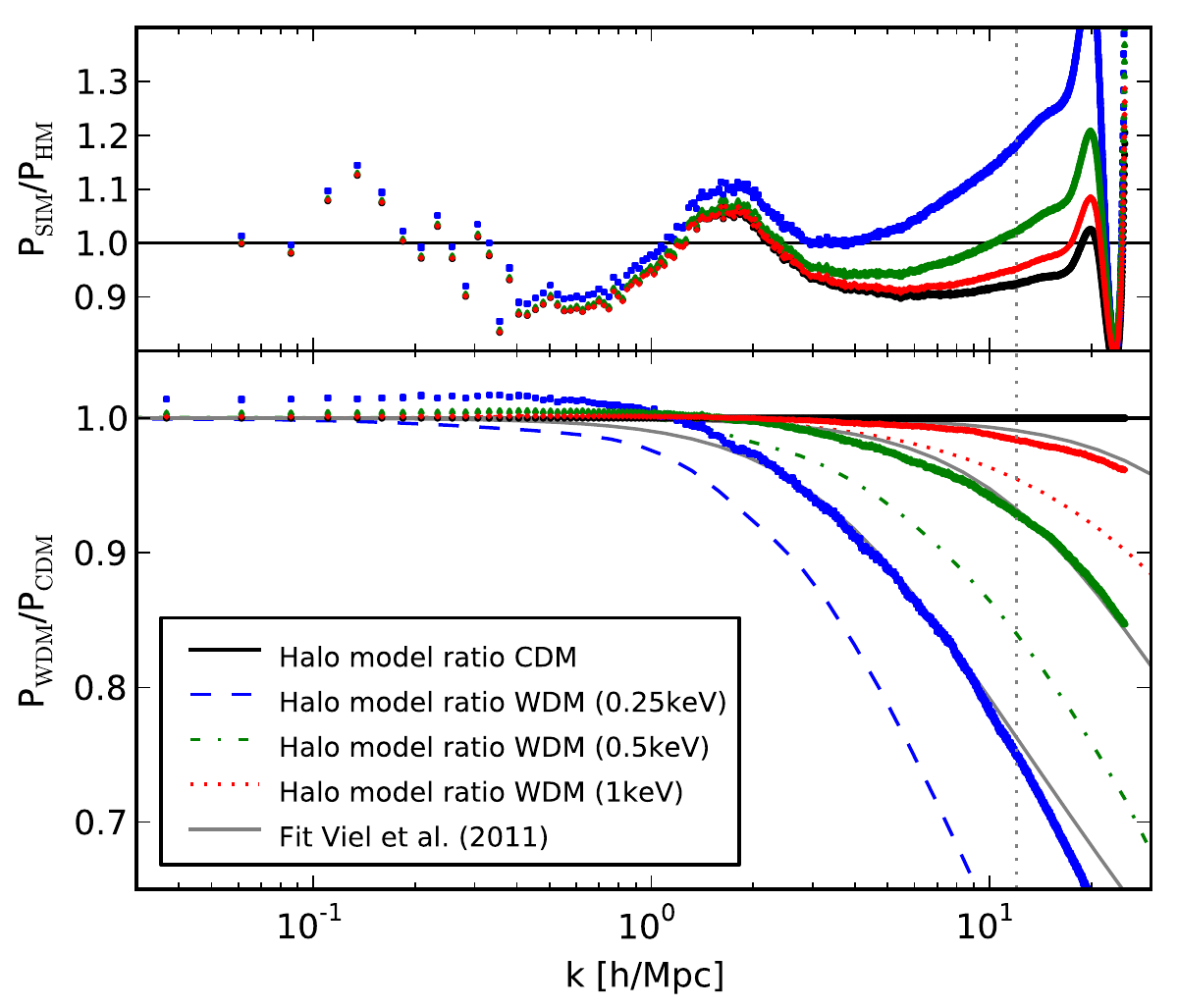}}
  \caption{\small{{\em Top panel}: Ratio of the simulated matter power
      spectra with respect to the halo model predictions as a function
      of wavenumber. Different coloured symbols denote the CDM and a
      selection of the WDM models. {\em Bottom panel:} Ratio of the
      WDM and CDM power spectra as a function of wavenumber.  Points
      denote the results from the ratios of simulation data; lines
      denote the halo model results. The gray solid lines correspond the
      fitting function from \citet{Vieletal2011}.}\label{PSratio0}}
\end{figure}


\section{Nonlinear power spectrum}\label{sec:Comparison}

\subsection{Comparison with existing models}

In \Fig{PScomparison0} we show the nonlinear matter power
spectra estimated from our highest resolution CDM and WDM
simulations. One can see that for $k\le1\kMpc$, there appears to be no
obvious difference between the CDM and WDM models under
consideration. This is in stark contrast with the initial linear theory
power spectra (cf.~\Fig{linPowSpec}), which show considerable damping
for the same scales. Clearly nonlinear evolution has regenerated a
high-$k$ tail to the power spectrum \citep[cf.][]{WhiteCroft2000}. At
higher wavenumbers $k>1\kMpc$, the situation is more interesting, and
we see that the measured WDM power spectra are suppressed with respect
to the CDM spectrum. The bottom panel of \Fig{PSratio0}
quantifies this suppression in greater detail. Here we see that at
$k\sim10\kMpc$ there is a 20\% suppression in power for the case of
$m_{\rm WDM}=0.25\keV$ and this drops to $\sim2\%$ for the case
$m_{\rm WDM}=1\keV$. The small difference between CDM and WDM at large scales ($k\lesssim 1$) is coming from a shift in the amplitude of the linear power spectrum, fixed with the same $\sigma_8$.

We now explore whether the halo model approach, described in
\S\ref{ssec:halomodel}, can accurately reproduce our results for the
WDM power spectra. In the original WDM halo model calculation of
\citet{SmithMarkovic2011}, all of the model ingredients (mass
function, density profiles and halo bias relation), were obtained by
assuming that the CDM relations also applied to the WDM case, provided
one computes them using the appropriate linear power spectrum.  The
results of this approach are presented in
\Figs{PScomparison0}{PSratio0} as the coloured line styles.


In \Fig{PSratio0} we see that the halo model of
\citet{SmithMarkovic2011} under-predicts the WDM power spectra by
roughly $\sim10\%$. This is reasonably good, considering the
assumptions that went into the model. This discrepancy was also noted
in the study of \citet{Vieletal2011}. In the bottom panel of
\Fig{PSratio0} we have also compared our nonlinear power spectra with
the predictions from the fitting formula presented in
\citet{Vieletal2011}. For scales $k<10\kMpc$ we find that this fitting
function provides an excellent description of our data. However, for
$k>10\kMpc$ we find discrepancies, especially for the case $m_{\rm
  WDM}=0.25\keV$. Whether this is a genuine failing of the fitting
formula is not clear, since this scale coincides with $\sim k_{\rm
  Ny}/2$, where $k_{\rm Ny}=\pi N_{\rm grid}/L$ is the Nyquist
frequency and we have used $N_{\rm grid}=2048$.
 
In summary, we find that the original halo model overestimates the
suppression of power due to WDM. This is not too surprising, since we
have seen in the previous section that the original approximations for
the halo mass function and concentrations turn out to be insufficient 
descriptions of the simulation data.


\begin{figure}
  \centering{
    \includegraphics[width=8.55cm]{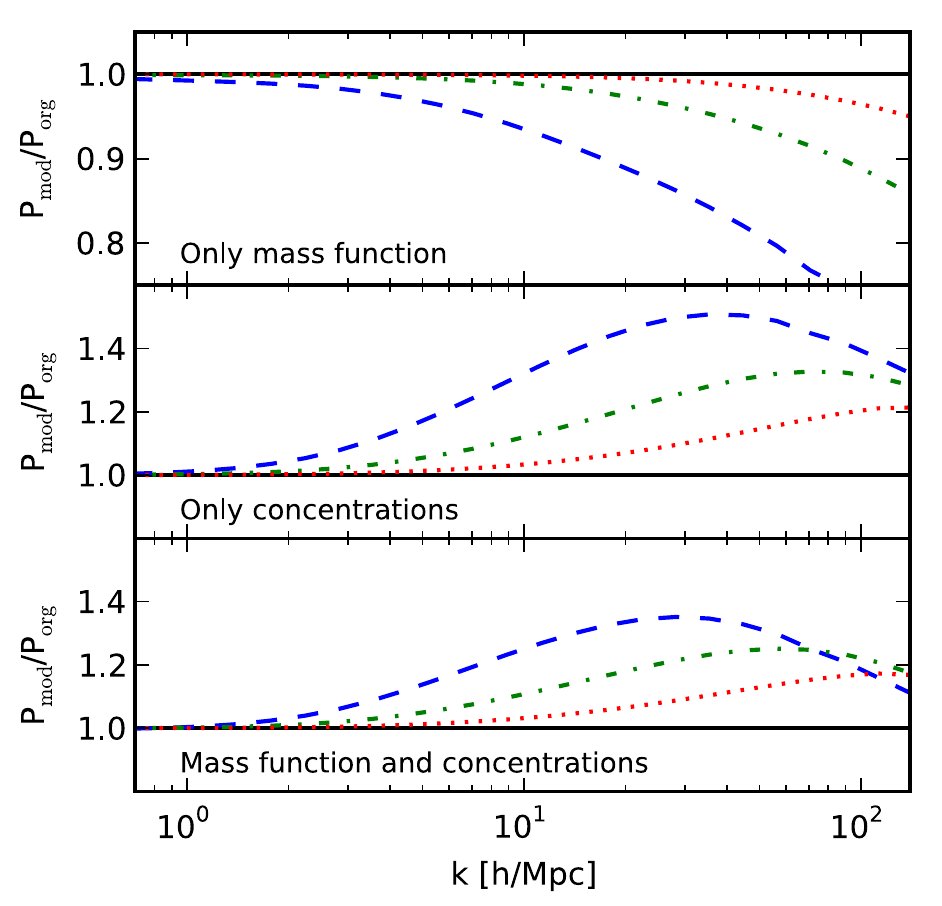}}
  \caption{\small{Ratio between modified versions of the halo model
      and the original version from \citet{SmithMarkovic2011}. The
      black solid line corresponds to CDM and the colored lines to WDM
      (red dotted: 1 keV, green dashed-dotted: 0.5 keV and blue
      dashed; 0.25 keV). Top panel: only modification of the mass
      function. Middle panel: only modification of the
      concentration-mass relation. Bottom panel: modification of mass
      function and concentration-mass relation.}\label{PSratios}}
\end{figure}


\subsection{Towards an improved WDM halo model}

We now explore whether the halo model predictions can be improved by
employing our better fitting functions for $n(M)$ and
$\rho(r|M)$. Before making a final prediction for $P(k)$, we first
examine how each modification affects the predictions individually.

If we implement our correction for the WDM mass function in the $P(k)$
predictions, then, since the abundance of small haloes is additionally
suppressed with respect to the predictions of $n^{\rm ST}$ for WDM, we
should expect that there is an even stronger suppression in $P(k)$.
This conjecture is confirmed in the top panel of \Fig{PSratios}, which
presents the ratio between the halo model with our modified mass
function and the original one. We clearly see that the ratio always
remains below unity. Somewhat surprisingly, we also note that a
$\sim50\%$ change in the abundance of $10^{12}\Msol$ haloes, leads to
a relatively small change, $\lesssim10\%$, in the power spectrum at
$k\lesssim10\kMpc$.

Next, if we instead implement our improved $c_{\rm vir}(M)$ relation,
then we find that this has a more significant impact on the spectra.
The central panel of \Fig{PSratios} shows the ratio between the
halo model with the modified concentrations and the original one.  We
find that the suppression of the halo concentrations leads to a
$\sim50\%$ boost in the power for $k\sim40\kMpc$.

The lower panel of \Fig{PSratios} shows the combined behavior of both
corrections. The ratio between the fully modified halo model and the
original one remains larger than unity. Thus, combination of the
modified $n(M)$ and $c(M)$, leads to halo model predictions that have
relatively more small-scale power.


\begin{figure}
  \centering{
    \includegraphics[width=8.5cm]{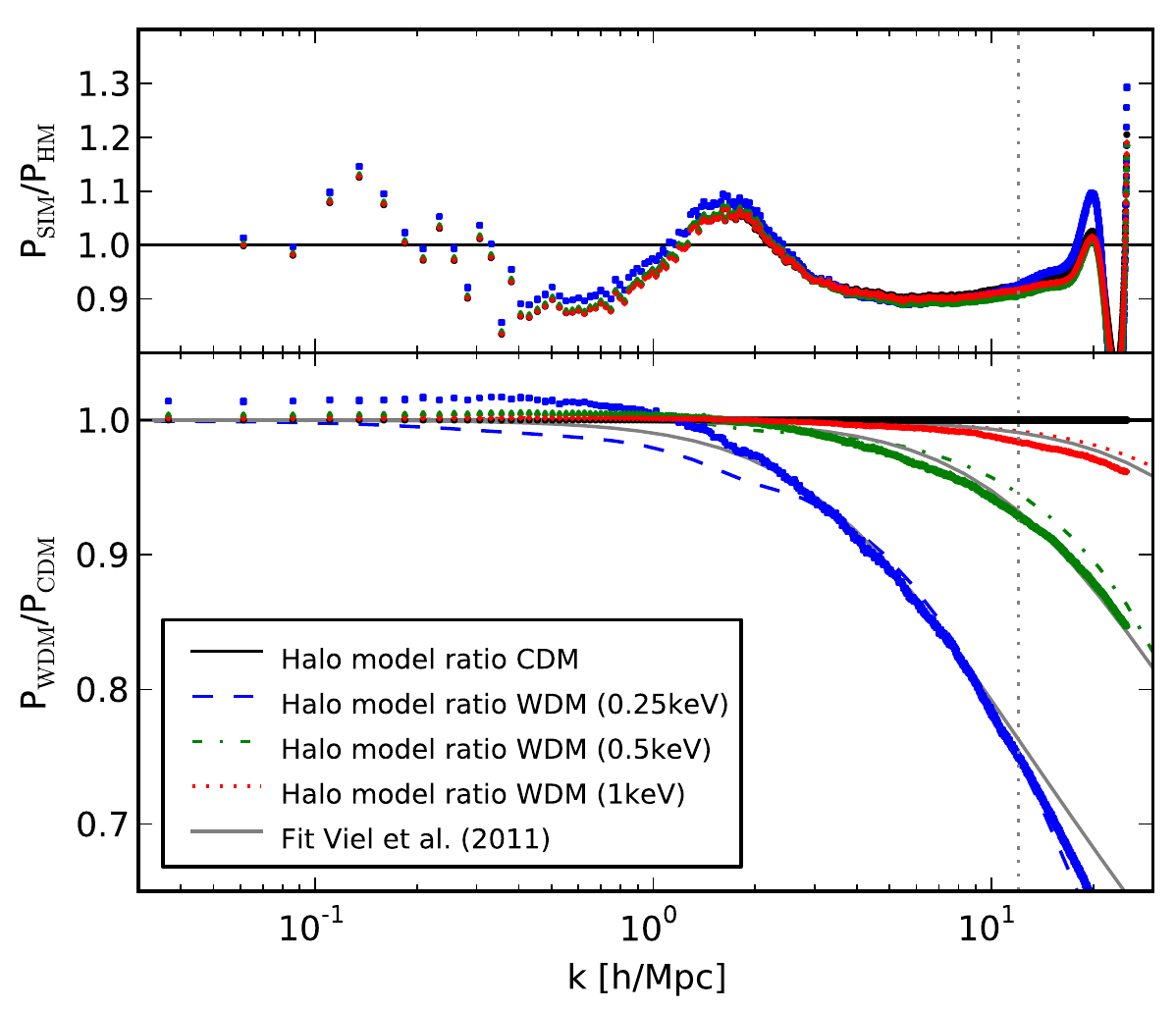}}
  \caption{\small{Nonlinear power spectra from the simulations (dots)
      and from the fully corrected halo model (lines), including the
      fits for the mass function and the concentrations. The labeling
      is the same than in \Fig{PScomparison0}. The error of the halo model compared to the simulations has dropped below 10 percent (top pannel), the error on the ratio between WDM and CDM has dropped well below 5 percent (bottom pannel).}
      \label{PScomparison1}}
\end{figure}


Finally, in \Fig{PScomparison1} we present the comparison
between our improved halo model and the nonlinear power spectra from
the simulations. The top panel presents the ratios between the
simulation data and the halo model predictions. The bottom panel shows
the ratios of the WDM and CDM results for both the simulations and our
modified halo model. Considering $k\gtrsim3\kMpc$, whilst our modified
halo model still has some problems predicting the overall absolute
value of $P(k)$, the relative changes between the WDM models and CDM
are almost exactly predicted, they being accurate to better than
$\sim2\%$ down to $k\sim10\kMpc$. For scales beyond $k\gtrsim10\kMpc$
we see that the halo model also matches the simulations very
well. However, again we note that these scales are beyond $k_{\rm
  Ny}/2$ and so one might worry about aliasing effects. For $k\lesssim 3$ the error is below about $5\%$, this scales are however suffering from the difficulties in calculating $P_{\rm hh}^{\rm c}$, descirbed in \S\ref{ssec:halomodel}.

In summary, we conclude that our modified halo model is able to
reproduce nonlinear WDM power spectra with the same accuracy as can
currently be achieved for CDM.


\section{Conclusion}\label{sec:Conclusion}

In this paper we have explored nonlinear structure formation in the
WDM cosmological model, through a large suite of cosmological $N$-body
simulations and through the halo model. The study was done for a set
of fully thermalized WDM models with particle masses in the set
$m=\{0.25,0.5,0.75,1.0,1.25\} \keV$. These masses range from purely
pedagogical models, towards more realistic scenarios for the dark
particle.

For the simulations we chose a box size ${L=256\Mpc}$, which was small
enough to resolve both the small scales, where WDM effects play an
important role, and the large scales, which are required for correct
linear evolution of the box-modes. All models were simulated with
$N=\{256^3, 512^3, 1024^3\}$ particles. This was done in order to
disentangle physical effects from numerical ones.

In the original halo model calculation for WDM by
\citet{SmithMarkovic2011}, it was shown that in order to make robust
predictions, one requires good understanding of dark matter halo
profiles, the mass function and halo bias.  In this work we performed
a detailed study of all of these ingredients. Our findings can be
summarized as follows:

\begin{enumerate}{\leftmargin = 1.0em}
\item Mass function: Below a certain scale, the WDM mass function is
  suppressed with respect to CDM. This suppression is considerably
  stronger than that obtained by simply applying the Sheth-Tormen
  approach together with the linear power spectrum of WDM. In
  agreement with \citet{SmithMarkovic2011}, we found that the mass
  functions for the different WDM models could be transformed into a
  single locus of points. This was achieved by taking the ratio of the
  WDM mass function with that for CDM, and then rescaling the masses
  by $M_{\rm hm}$ (or equivalently $M_{\rm fs}$). We used a fitting
  function similar to that proposed in \citet{Dunstanetal2011} to link
  the Sheth-Tormen mass function to the measured one. The fitting
  function, which has only one free parameter, was able to reproduce
  all of the data with an accuracy of a few percent. We also found a
  strong boost in the mass function at very small mass scales. We
  showed that this was consistent with artificial halo formation
  around the initial particle lattice \citep[cf.][]{WangWhite2007}.

\item Halo bias: We measured the linear halo bias, using the four
  largest modes in our simulations. For smaller mass haloes, we found
  a small enhancement of the bias in WDM simulations, which was
  qualitatively consistent with the predictions of
  \citet{SmithMarkovic2011}. However, owing to the simulation box
  being too small, we were unable to quantify this more robustly.  At
  very small masses we found a prominent boost in the bias. We found
  that this was again a sign of artificial halo formation.

\item Density profiles: In the CDM model, the density profiles of dark
  matter haloes can be characterized by an NFW profile, with a
  monotonically decreasing concentration-mass relation. In the WDM
  scenario, we have shown that the NFW profile remains valid for the
  models and resolution limits of our simulations, and we saw no
  evidence for a central density core. A simple adaption of the CDM
  concentration-mass relation, would suggest a strong flattening
  towards small masses. Whilst, we found such a flattening, the
  measurements in fact revealed a turnover towards smaller
  masses. This somewhat surprising result may be interpreted as a sign
  of top-down structure formation.  We modelled the mean relation by
  adapting a fitting formula similar to that for the mass
  function. Our fit to the $c(M)$ data was good to an accuracy of
  $\sim10\%$. Interestingly, we found that the deviations from CDM in
  the WDM model, appear in the $c(M)$ relation for halo masses one
  order of magnitude larger than for the mass function.

\end{enumerate}

After analyzing these ingredients in detail and developing new fitting
functions for them, we were able to improve the small-scale
performance of the WDM halo model. We found that for $k\gtrsim3\kMpc$,
we could predict the absolute amplitude of the power spectrum to
better than $\sim10\%$. However, we were able to predict the ratio of
the WDM to CDM spectra, at better than $\lesssim2\%$. This was
competitive with the latest fitting formulae \citep{Vieletal2011}.

One of the many advantages of the halo model based approach, is that
we may more confidently extrapolate our power spectra predictions to
smaller scales than can be done from a fitting formula, since the
model is built on physical quantities. Furthermore, we may also use
the model to study the clustering of galaxies
\citep{Zehavietal2005}. It is hoped that this may lead to a method for
constraining WDM models from galaxy clustering studies.  Lastly, one
further issue for future study, is to establish a better theoretical
understanding of what shapes the mass function and halo concentrations
in WDM. In particular, in finding the turnover in the concentration
mass relation, have we really seen the reversal of bottom-up structure
formation. This promises to be an interesting future challenge.


\section*{Acknowledgments}

It is a pleasure to thank Donnino Anderhalden, J\"urg Diemand and
Darren Reed for useful discussions. We also thank Doug Potter and
Joachim Stadel for use of their power spectrum code and technical
support concerning {\tt pkdgrav}. We thank Roman Scoccimarro for
making public his {\tt 2LPT} code. AS, RES and BM acknowledge support
from the Swiss National Foundation. RES also acknowledges support from
a Marie Curie Reintegration Grant and the Alexander von Humboldt
Foundation. 




\input{mnras1.bbl}

\end{document}

%% file: defs.tex
\newcommand\Eqn[1]     {Eq.\,(\ref{#1})}
\newcommand\Eqns[2]    {Eqs\,(\ref{#1}) and~(\ref{#2})}

\newcommand\Fig[1]     {Fig.\,{\ref{#1}}}
\newcommand\Figs[2]    {Figs\,{\ref{#1}} and~{\ref{#2}}}

\def\mnras{{Mon.~ Not.~ R.~ Astron.~ Soc.~}}

\def\prd{{Phys.~ Rev.~ D.~}}

\def\apj{{Astrophys.~ J.~}}
\def\apjs{{Astrophys.~ J.~ Suppl.~}}
\def\apjl{{Astrophys.~ J.~ Lett.~}}

\def\aj{{Astron.~ Journal}}
\def\nat{{Nature (London)~}}

\def\mnras{{MNRAS}}

\def\prd{{PRD}}

\def\apj{{ApJ}}
\def\apjs{{ApJS}}
\def\apjl{{ApJL}}

\def\nat{{Nature}}

\def\physrep{{Phys.~ Rep.~}}
\def\jcap{Journal of Cosmology and Astro-Particle Physics}

\newcommand{\be}{\begin{equation}}
\newcommand{\ee}{\end{equation}}
\newcommand{\ba}{\begin{eqnarray}}
\newcommand{\ea}{\end{eqnarray}}

\def\pp1{{\prime}}
\def\pp2{{\prime\prime}}

\def\2D{{\rm 2D}}

\def\h{{\rm h}}

\def\bx{{\bf x}}

\def\1Loop{{\rm 1Loop}}

\def\rhob{\bar{\rho}}

\def\Msol{h^{-1}M_{\odot}}
\def\kpc{\, h^{-1}{\rm kpc}}
\def\Mpc{\, h^{-1}{\rm Mpc}}

\def\kMpc{\, h \, {\rm Mpc}^{-1}}

\def\dk{{\rm d}^3{\bf k}}

\font\BF=cmmib10

\def\v{{\hbox{\BF v}}}

\def\la{\mathrel{\mathpalette\fun <}}
\def\ga{\mathrel{\mathpalette\fun >}}
\def\fun#1#2{\lower3.6pt\vbox{\baselineskip0pt\lineskip.9pt
        \ialign{$\mathsurround=0pt#1\hfill##\hfil$\crcr#2\crcr\sim\crcr}}}

\def\s{{\rm s}}
\def\h{{\rm h}}

\def\1H{{\rm 1H}}
\def\2H{{\rm 2H}}

\def\WDM{{\rm WDM}}

\def\keV{{\rm keV}}


%% file: images.tex
\begin{figure*}
  \centering{
    \includegraphics[width=5.8cm]{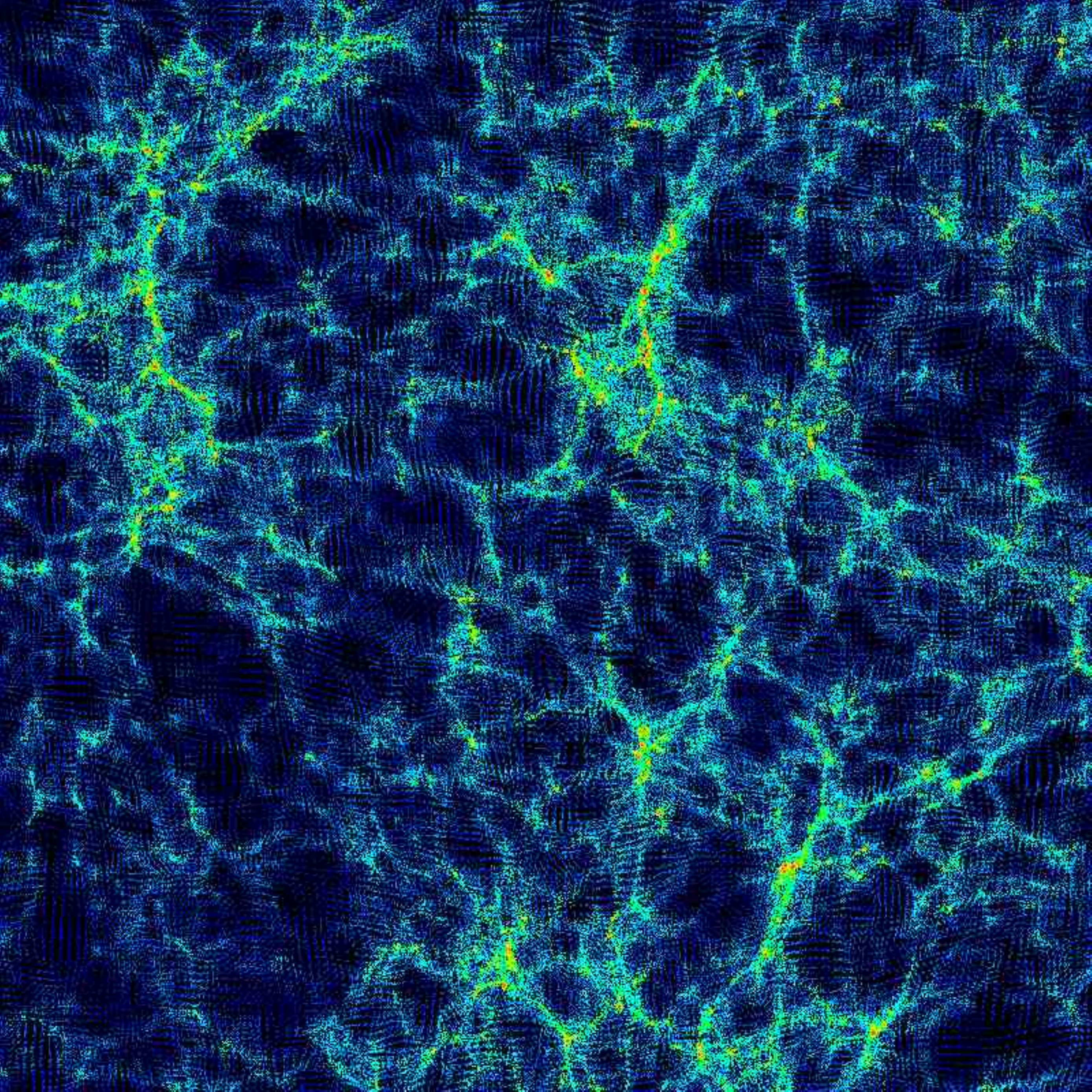} 
    \includegraphics[width=5.8cm]{IMAGES/CDM_L256_N1024_00010RRv6.pdf}
    \includegraphics[width=5.8cm]{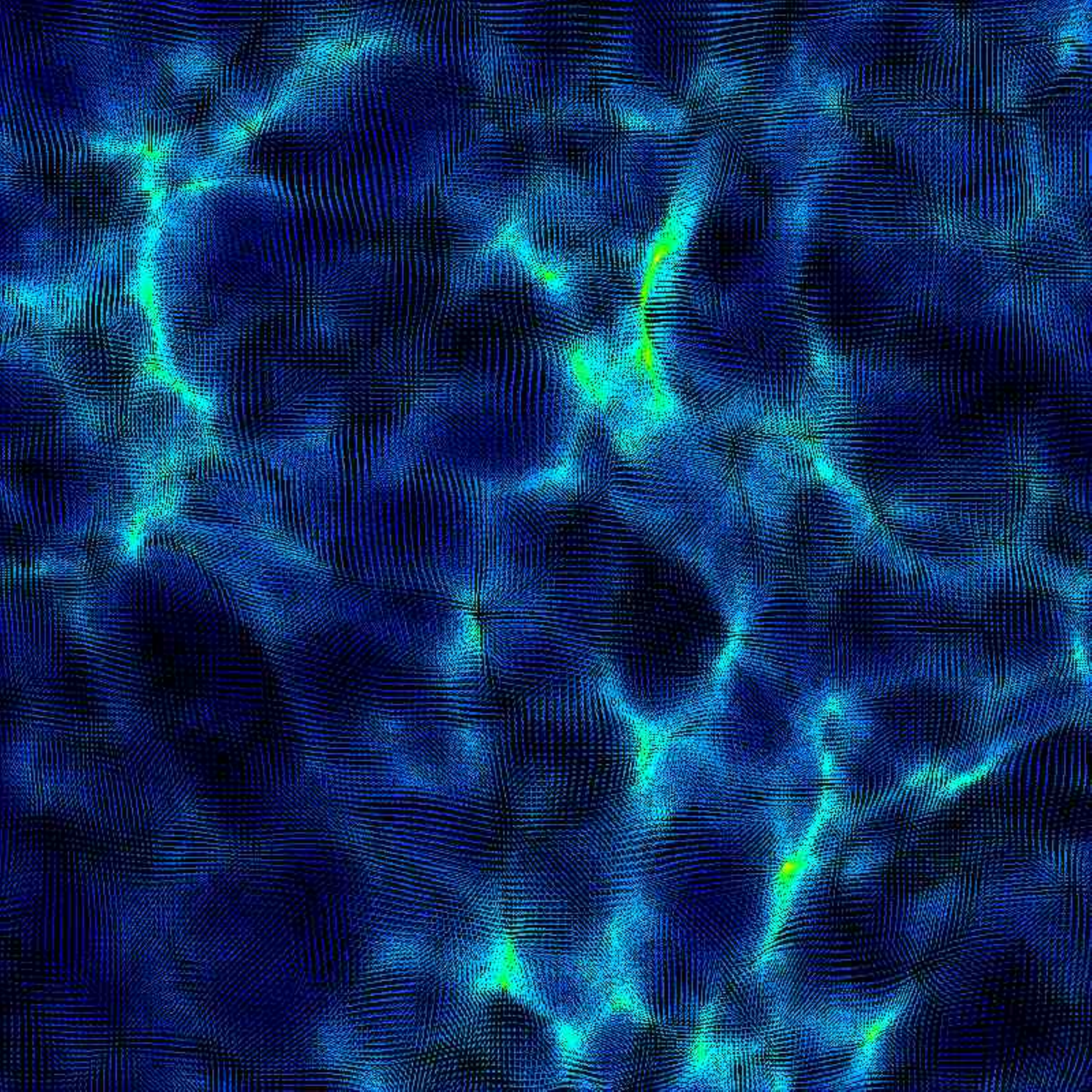}}
  \vspace{0.05cm}
  \centering{
    \includegraphics[width=5.8cm]{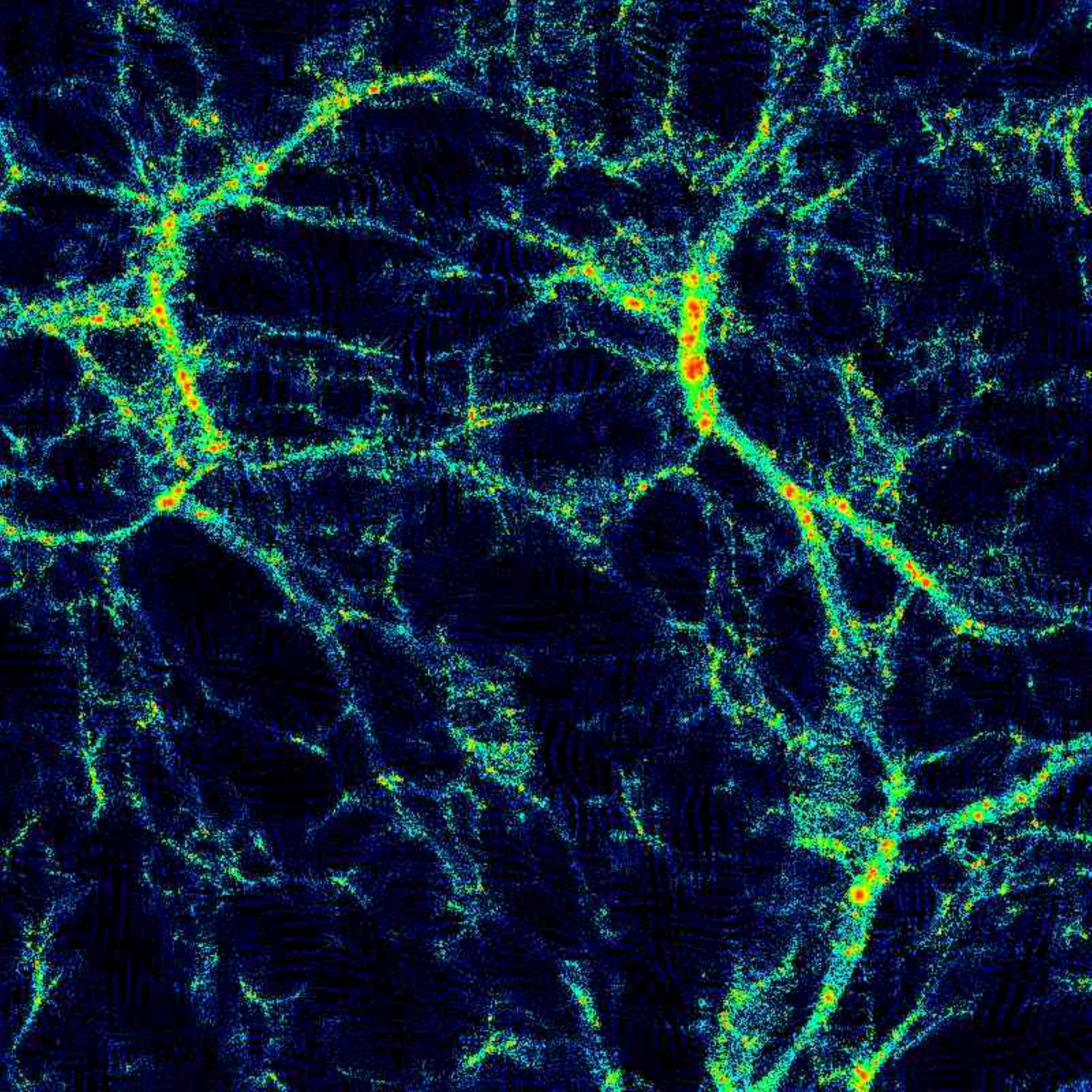}
    \includegraphics[width=5.8cm]{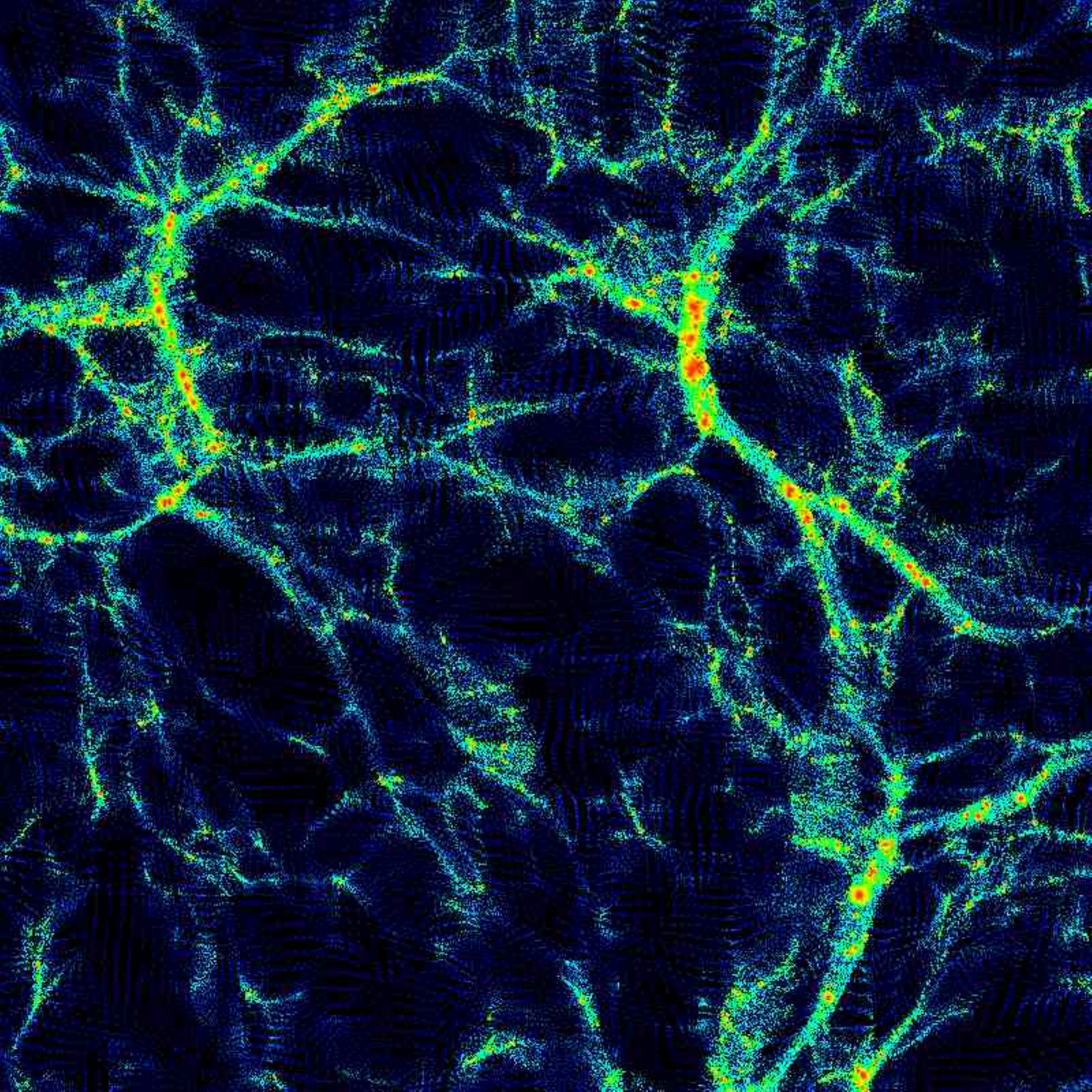}
    \includegraphics[width=5.8cm]{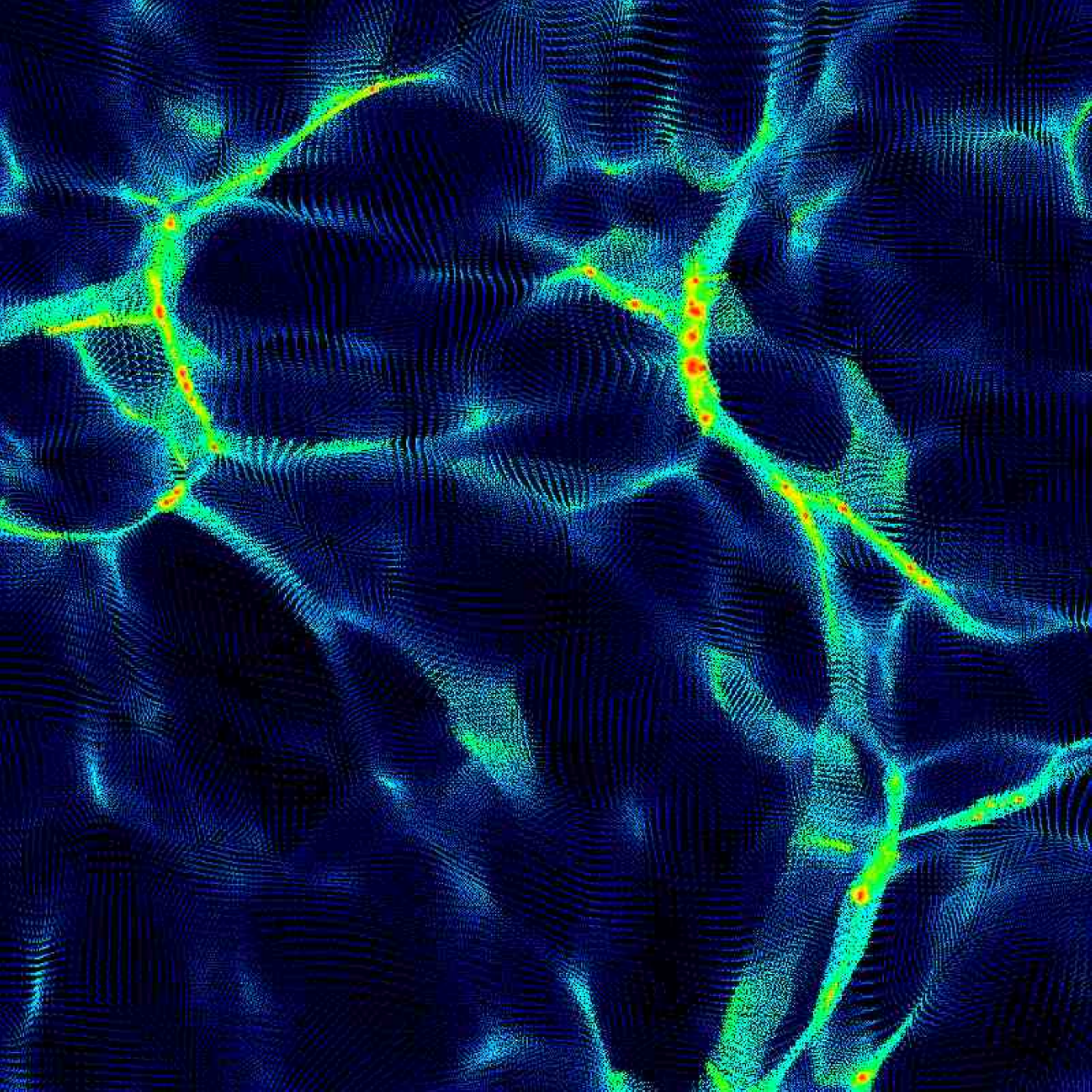}}
  \vspace{0.05cm}
  \centering{
    \includegraphics[width=5.8cm]{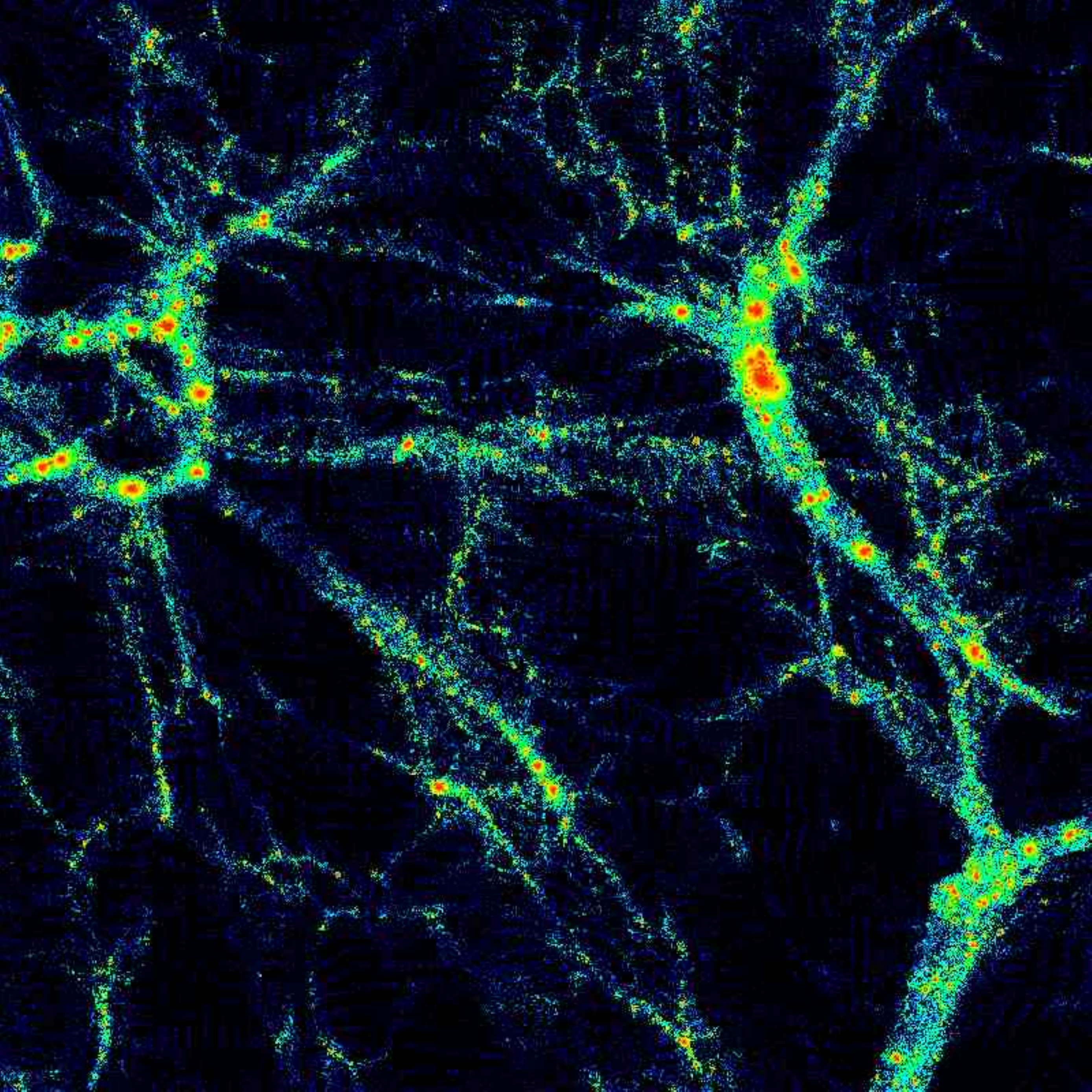}
    \includegraphics[width=5.8cm]{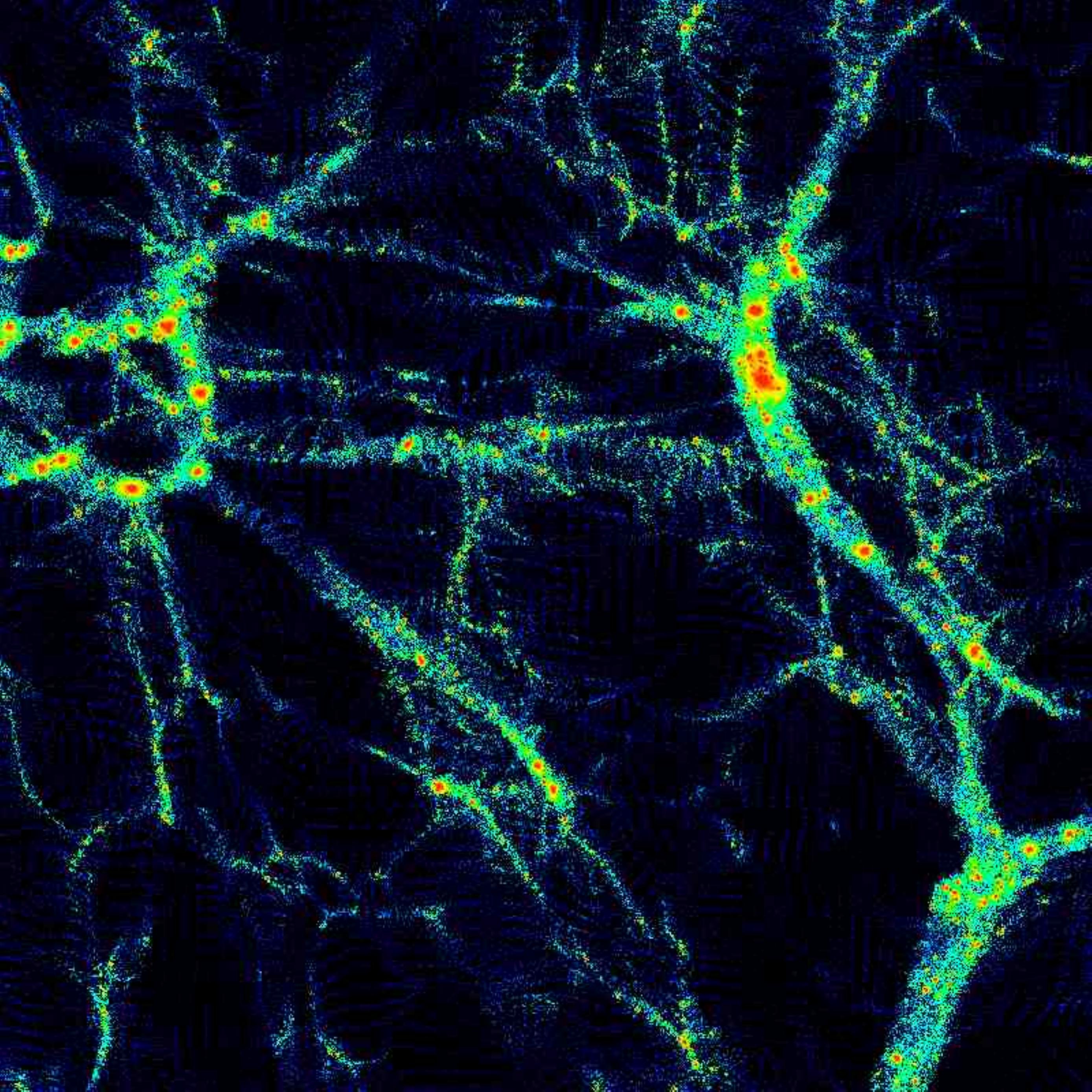}
    \includegraphics[width=5.8cm]{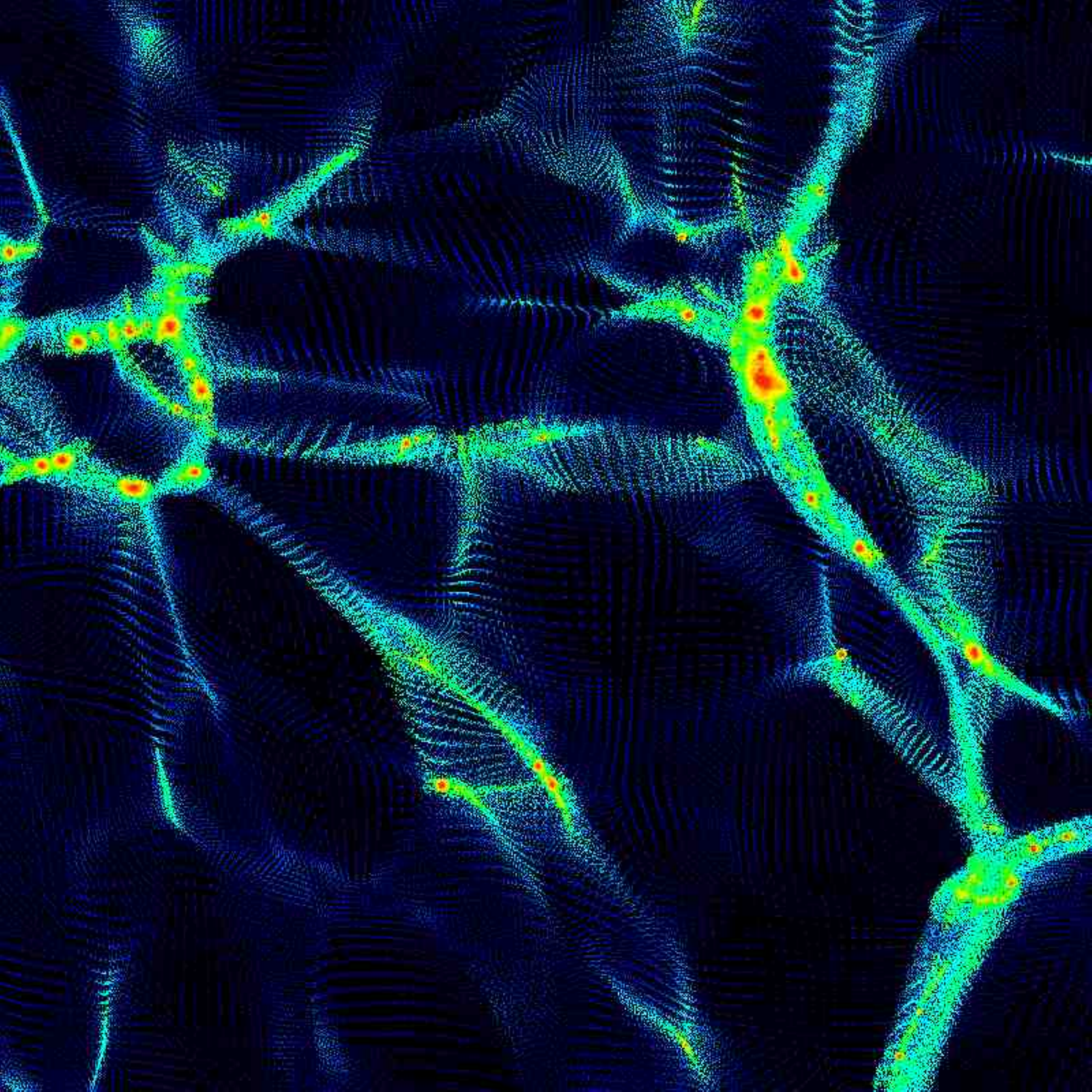}}
  \vspace{0.05cm}
  \caption{\small{Density maps from the $N=1024$ simulations with a length of
      $50\Mpc$ and a depth of $2.5\Mpc$. From top to bottom: $z=4.4$, $z=1.1$ and $z=0$. From
      left to right: CDM, WDM with $m_{\rm p}=1.0\keV$ and WDM with
      $m_{\rm p}=0.25\keV$.} Whilst the WDM effects are barely
      discernible in the middle panels, they are very prominent in the
      right panels, where the voids are noticeably emptier than in
      CDM.\label{SimImages}}
\end{figure*}